\documentclass[%
 reprint,
superscriptaddress,
 amsmath,amssymb,
 aps,
]{revtex4-2}

\usepackage{graphicx}
\usepackage{dcolumn}
\usepackage{bm}

\usepackage{gensymb}
\begin{document}

\title{Move aside pentacene: Diazapentacene doped para-terphenyl as a zero-field room-temperature maser with strong coupling for cavity quantum electrodynamics}

\author{Wern Ng}
\email{wern.ng@imperial.ac.uk}
 \affiliation{Department of Materials, Imperial College London, South Kensington, SW7 2AZ London, United Kingdom}

 \author{Xiaotian Xu}
 \affiliation{Department of Materials, Imperial College London, South Kensington, SW7 2AZ London, United Kingdom}
 
  \author{Max Attwood}
 \affiliation{Department of Materials, Imperial College London, South Kensington, SW7 2AZ London, United Kingdom}
 
  \author{Hao Wu}
 \affiliation{Center for Quantum Technology Research and Key Laboratory of Advanced Optoelectronic Quantum Architecture and Measurements (MOE), School of Physics, Beijing Institute of Technology, Beijing 100081, China}
  \affiliation{Beijing Academy of Quantum Information Sciences, Beijing 100193, China}
 
 \author{Zhu Meng}
 \affiliation{Department of Chemistry and Centre for Processible Electronics, Imperial College London, W12 0BZ London, United Kingdom}

 \author{Xi Chen}
 \affiliation{Department of Materials, Imperial College London, South Kensington, SW7 2AZ London, United Kingdom}
 \affiliation{Department of Computer Science, University of Southern California,  Los Angeles, California, USA}

 \author{Mark Oxborrow}
 \altaffiliation{m.oxborrow@imperial.ac.uk}
 \affiliation{Department of Materials, Imperial College London, South Kensington, SW7 2AZ London, United Kingdom}

\date{\today}

\begin{abstract}
    Masers, the microwave analogue of lasers, promise to deliver ultra-low noise amplification of microwave signals for use in medical MRI imaging and deep-space communication. Research on masers in modern times was rekindled thanks to the discovery of gain media that were operable at room-temperature, eschewing bulky cryogenics that hindered their use. However, besides the two known materials of pentacene doped in para-terphenyl and negatively-charged nitrogen-vacancy defects in diamond, there has been scarce progress in the search for completely new room-temperature gain media. Here we show the discovery of 6,13-diazapentacene doped in para-terphenyl as a maser gain medium that can operate at room-temperature and without an external magnetic field. A measured maser pulse power of -10 dBm shows it is on par with pentacene-doped para-terphenyl in absolute power, while possessing compelling advantages against its pentacene predecessor in that it has a faster amplification startup time, can be excited with longer wavelength light at 620 nm and enjoys greater chemical stability from added nitrogen groups. Furthermore, we show that the maser bursts allow 6,13-diazapentacene-doped para-terphenyl to reach the strong coupling regime for cavity quantum electrodynamics, where it has a high cooperativity of 182. We study the optical and microwave spin dynamics of 6,13-diazapentacene-doped para-terphenyl in order to evaluate its behavior as a maser gain medium, where it features fast intersystem crossing and an advantageously higher triplet quantum yield. Our results pave the way for the future discovery of other similar maser materials and help point to such materials as promising candidates for the study of cavity quantum electrodynamic effects at room-temperature.
\end{abstract}

\maketitle
Accessing the strong-coupling regime of cavity quantum electrodynamics (cQED) in a benchtop system operating at room-temperature represents an alluring engineering goal. A high quality-factor ($Q$) microwave resonator magnetically loaded with optically-pumped pentacene doped in para-terphenyl (Pc:PTP) provides one route towards this goal \cite{Breeze2017cqed}, and diamond with negatively-charged nitrogen-vacancy defects (NV$^-$ diamond) has also been considered\cite{Zhang2022}. However attaining a sufficiently high spin polarisation density for strong coupling from more materials has, at room-temperature, proven elusive. 
\begin{figure*}[htbp]
  \centering
  \includegraphics[width=0.9\linewidth]{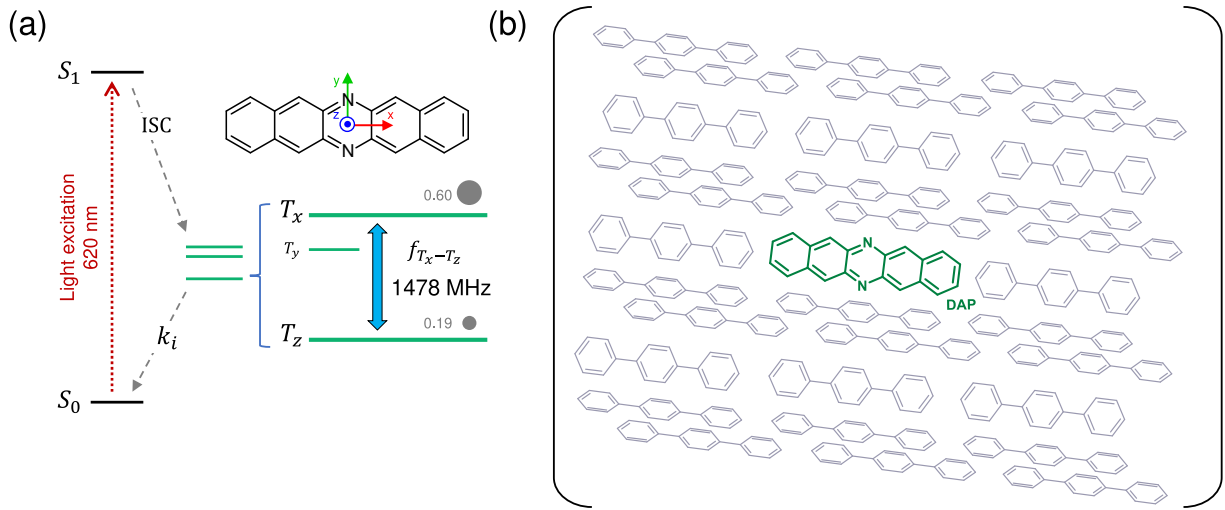}
  \caption{(a) Simplified Jablonski diagram showing optical pumping of DAP:PTP using 620 nm light to the singlet excited state. The population of spins then transfer to the triplet sublevels (each labelled according to the molecular axes drawn on a DAP molecule) through intersystem crossing, where the frequency of the transition between $T_x$ and $T_z$ is shown and the sublevel population ratios are given. The populations then decay back to the singlet ground state through depopulating rates labelled as $k_i$, where $i=x,y,z$ refers to different depopulating rates for each sublevel. (b) Drawing of a DAP molecule (green) doped into a unit cell of a PTP lattice (grey), where the DAP substitutes out one of the PTP molecules.}
  \label{fig:simplejablon}
\end{figure*}
Room-temperature masers (microwave amplification by stimulated emission of radiation) can be made out of either Pc:PTP or NV$^-$ diamond\cite{Oxborrow2012,Breeze2018}, but Pc:PTP remains the sole material able to mase without any externally applied magnetic field (zero-field) at room-temperature. However, it has a drawback in that it suffers from lag (of several microseconds) at start up; when operating in pulsed mode (e.g. providing low-noise amplification solely when required) this lag excludes applications that require a rapid response. Since masers on the benchtop remain as tantalising prospects for achieving ultra-low noise amplification applicable to MRI in medical devices, deep space communication and for demonstrating cQED at room-temperature\cite{Breeze2017cqed}, the discovery of other maser materials could help accelerate developments in these fields.

Previous literature had investigated the electronic structures and spin dynamics of various molecular alternatives to Pc:PTP as masers\cite{Bogatko2016}. A particularly promising candidate was 6,13-diazapentacene doped in para-terphenyl (DAP:PTP), which was postulated to mase through a population inversion between its photoexcited triplet sublevels, $T_x$ and $T_z$\cite{Bogatko2016}. However, no attempt was made to make a maser out of it or characterise the spin dynamics of its triplet state. In this work, we report on the successful masing of DAP:PTP on its $T_x-T_z$ transition near 1478 MHz, as only the second gain material discovered that can do so at both zero-field and room-temperature. The steady-state and transient optical absorption properties of DAP:PTP were characterised alongside its triplet state dynamics using zero-field transient electron paramagnetic resonance (ZF-trEPR). The experimental masing signals were then measured, which showcased Rabi oscillations that were simulated to show the high cooperativity of DAP:PTP in the strong coupling regime. This all highlights DAP:PTP as an advantageous material for not only masers but also the exploration of cQED at room-temperature.
\section{Results}

\subsection{Optical Characterisation}

\begin{figure*}[htbp]
  \includegraphics[width=\linewidth]{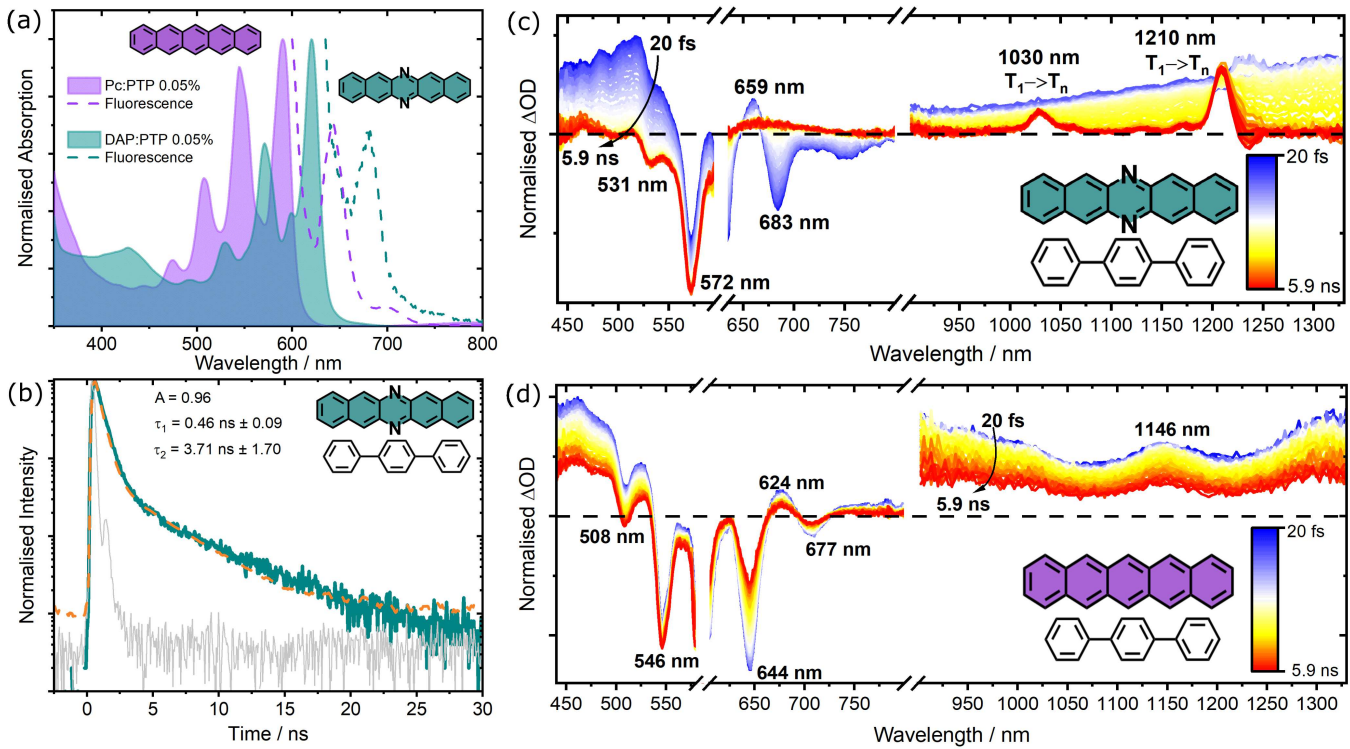}
  \caption{Optical characterisation of DAP:PTP and Pc:PTP. (a) Steady-state UV/Vis absorption and fluorescence spectrum after excitation at 590 nm for Pc:PTP and 620 nm for DAP:PTP; (b) TCSPC spectroscopy of 0.05\% DAP:PTP (teal curve, $\lambda_{ex}$ = 404 nm, $\lambda_{em}$ = 626 nm). Orange dashed lines represent the fit using a bi-exponential decay function, grey solid lines represent the instrument response function (IRF); (c) fsTAS of 0.05\% DAP:PTP with $\lambda_{pump}$ = 620 nm and (d) Pc:PTP with $\lambda_{pump}$ = 590 nm}
  \label{fig:CombinedOpticalProperties_Final}
\end{figure*}

DAP:PTP has a population inversion between the $T_x$ and $T_z$ sublevels\cite{Bogatko2016}, and the ratio in which these sublevels are populated depends on optical pumping from a singlet ground state ($S_0$) to a singlet excited state ($S_1$), followed by intersystem crossing (ISC) that selectively populates the sublevels differently. A simplified Jablonski diagram in Figure~\ref{fig:simplejablon}(a) illustrates this process (alongside a depiction of a DAP:PTP lattice in Figure~\ref{fig:simplejablon}(b)), and so the optical properties of singlet excitation rates, ISC and depopulation away from the triplet states all play a vital role in the masing capabilities of DAP:PTP. To begin interrogating the optical properties of DAP:PTP, we initially compared its steady-state visible optical properties to Pc:PTP. DAP:PTP exhibits a UV/Vis absorption spectrum and fluorescence output red-shifted by approximately 30 nm compared to Pc:PTP with characteristic Frank-Condon vibronic bands and $\lambda_{max}$ = 620 nm (Figure~\ref{fig:CombinedOpticalProperties_Final}(a)). The ability to pump maser gain media at lower optical frequencies enhances their potential efficiency, since the number of absorbed photons per unit of pump energy increases. Since the majority of absorbed pump energy is lost internally through vibrational relaxation and heating, reducing the pump energy also minimises the risk of charring the gain material. 

The transient excited state dynamics of DAP:PTP following photoexcitation are also important for judging its merit as a maser. We performed time-correlated single photon counting (TCSPC) spectroscopy on single crystal shards grown within 4 x 0.6 mm rectangular capillary tubes. To achieve sufficient signal-to-noise for the measurement, a higher-than-saturation concentration of 0.05\% DAP:PTP was achieved using a relatively rapid crystal growth rate of 10 mm/hr. However, this resulted in significantly smaller individual crystal domains than samples grown at a slower rate (4 mm/hr growth) for maser experiments.

The fluorescence lifetime ($\tau_{f}$) of DAP:PTP was found to decay bi-exponentially, with a mono-exponential fitting function unable to describe the earliest time points and a tri-exponential decay offering negligible improvement to the fitting (Figure~\ref{fig:CombinedOpticalProperties_Final}(b)). This analysis returned a relatively short lived decay lifetime of approximately 0.46 ns (A = 0.96) alongside a longer lifetime of 3.7 ns. By comparison, the $\tau_{f}$ of Pc:PTP is known to be approximately 9 ns at room-temperature\cite{Patterson1984}. This confirms that despite being based on the same molecular scaffold, the nitrogen substitution on 6,13-diazapentacene (DAP) results in significantly faster excited state dynamics. The necessity for a bi-exponential fitting suggests two dominant decay mechanisms from $S_{1}$. Typical relaxation routes for diluted solutions, films or crystals of linear acenes include fluorescence and non-radiative decay, the latter of which can consist of vibrational decay from $S_{1}$ to $S_{0}$ or ISC into the triplet manifold. At 0.05\% doping concentration, alternative decay mechanisms such as triplet-triplet annihilation are not expected to contribute significantly. DAP and pentacene (Pc) are similar in structure and extent of electron delocalisation, so their transition dipole moments and resulting rates of fluorescent relaxation are likely to be similar\cite{Wang2022}. Therefore, the reduced fluorescence lifetime of DAP:PTP compared to Pc:PTP indicates a faster rate of ISC. 

\begin{figure*}[htp]
  \includegraphics[width=\linewidth]{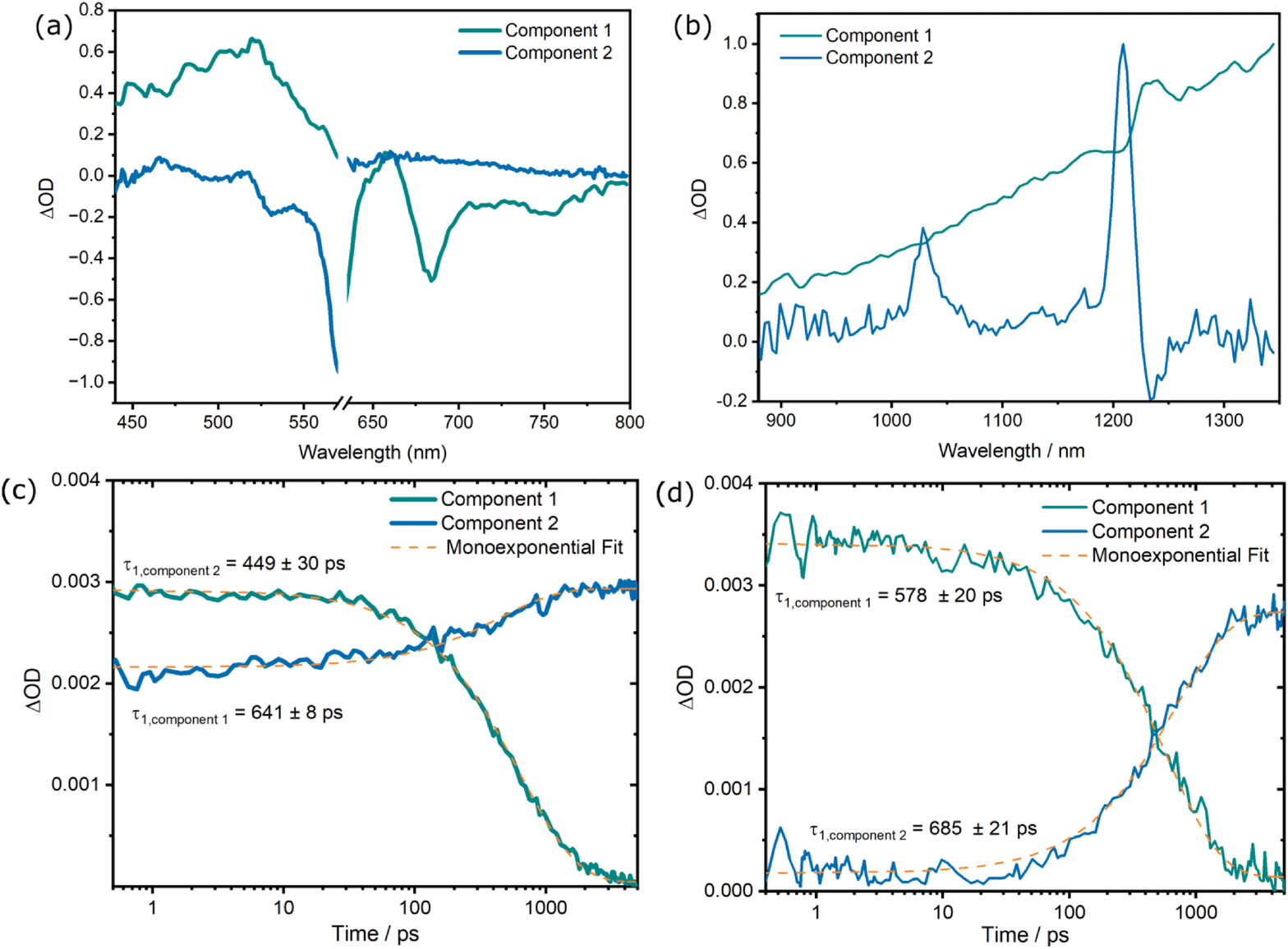}
  \caption{[Extended Data Figure] (a) Visible spectrum components for DAP:PTP fsTAS data; (b) Near-infrared (NIR) spectrum components; (c) visible component time profiles; (d) NIR component time profiles. Extracted decay time constants were determined using a mono-exponential fitting (orange dashed line).}
  \label{fig:dapfsTAS data}
\end{figure*}

To further assess the decay of $S_{1}$, visible and near-infrared (NIR) femtosecond transient absorption spectroscopy (fsTAS) was utilised to search for signature singlet and triplet state absorption bands. Measurements were conducted up to 5.9 ns, thereby covering the majority of $S_{1}$ decay as indicated by TCSPC spectroscopy. The TAS spectrum of DAP:PTP exhibited a negative ground state bleach (GSB) at 531 and 572 nm, corresponding to Frank-Condon absorption bands in the steady-state visible spectrum (Figure~\ref{fig:CombinedOpticalProperties_Final}(c)). The broad positive baseline between 440 and 550 nm is typical of excited state absorption (ESA) from $S_{1}$ to $S_{n}$ transitions, and is likely superimposed throughout most of the visible region. This is also evidenced in the NIR region with a growing baseline up to 1350 nm. At early time points, negative bands are also observed at approximately 630 and 683 nm which can be identified as fluorescence vibronic bands seen in the fluorescence spectrum of DAP:PTP. Together, these fluorescence and ESA features decay back to the baseline within 1200 ps, leaving in-growing GSB bands at 531 and 572 nm. The growth of the GSB indicates a depletion of the singlet state overall and is consistent with an increase in the presence of relatively long-lived triplet states. 
\begin{figure*}[htbp]
\centering
  \includegraphics[width=0.9\linewidth]{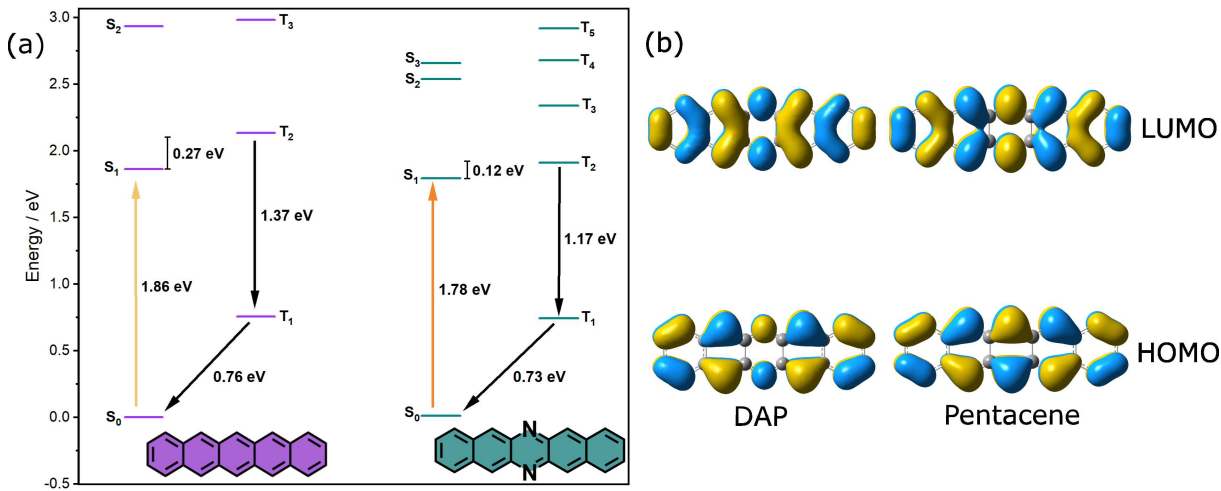}
  \caption{(a) Expanded Jablonski diagram for Pc and DAP constructed with TD-DFT-derived energies for the excited electronic states of singlet and triplet optimised geometries. (b) Calculated HOMO and LUMO orbitals of DAP and Pc.}
  \label{fig:DFT-derived State Energies}
\end{figure*}
Further evidence for the involvement of triplet states is observed in the NIR region, where two absorption bands are revealed as early as 200 ps at 1030 and 1210 nm, growing until termination of the experiment at 5.9 ns. Since in-growing spectral features are typical of the triplet state, we attribute these bands to ESA transitions between $T_{1}$ and $T_{n}$ states. Global analysis by singular value decomposition (SVD) of these spectra reveal two significant principle components for both visible and NIR regions, describing ESA and GSB features, respectively (Extended Figure~\ref{fig:dapfsTAS data}). A mono-exponential fit reveals lifetime values of 449 and 580 ps for the first component of the visible and NIR regions, respectively. These values are reasonably close to the 460 ps fluorescence lifetime determined via TCSPC spectroscopy. This strongly indicates that the first components describing the visible and NIR regions represent the total depletion of $S_{1}$ states. Since these components are a function of the abundance of $S_{n}$ states, their decay is intrinsically linked to both fluorescence and non-radiative decay, i.e. internal conversion  (IC) and ISC. However, triplet absorption is not, hence the time constant of 685 ns derived from the second component in the NIR region can be associated with the rate of ISC ($\kappa_{ISC}$ = 1/$\tau_{ISC}$), yielding a rate of 1.46 ns$^{-1}$. This is consistent with the earlier assertion that the in-growing GSB features, which exhibit a very similar 641 ps lifetime, are linked with an increase in the presence of the triplet state. Since the total rate of fluorescence  ($\kappa_{f}$ = 1/$\tau_{f}$ = 2.174 ns$^{-1}$) is the sum of the rates of radiative decay ($\kappa_{rad}$), IC ($\kappa_{IC}$) and ISC ($\kappa_{ISC}$), it can be determined that $\kappa_{IC}$ + $\kappa_{rad}$ = 0.714 ns$^{-1}$. Having quantified $\kappa_{f}$ and $\kappa_{ISC}$, the triplet quantum yield ($\theta_{T}$) can be estimated as their ratio\cite{deSouza2016}, yielding $\theta_{T}=67\%$, higher than the accepted $\theta_{T}= 62.5\%$ for Pc:PTP\cite{Takeda2002}.

\begin{figure*}[htp]
  \includegraphics[width=\linewidth]{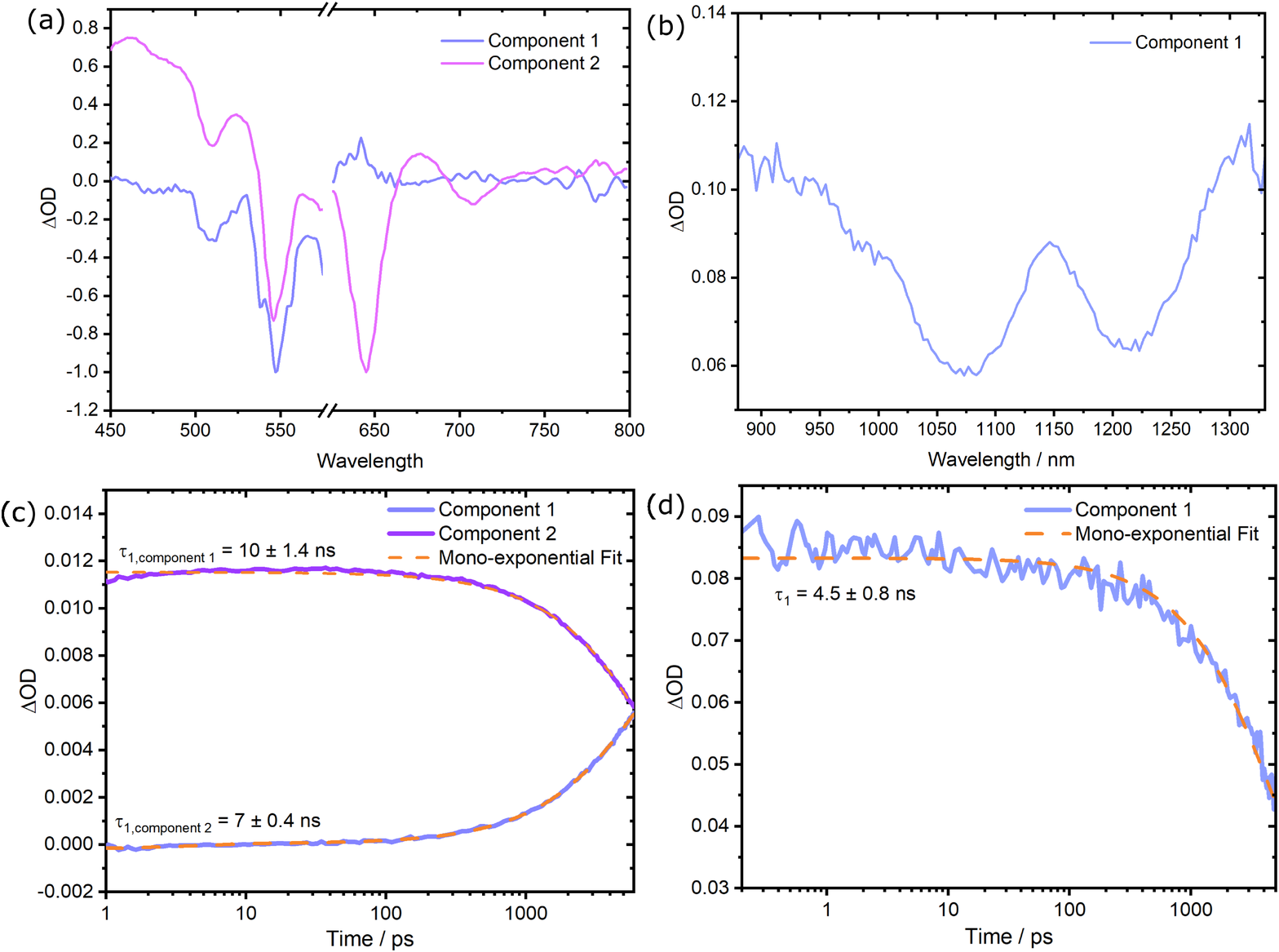}
  \caption{[Extended Data Figure] (a) Visible spectrum components for Pc:PTP fsTAS data; (b) NIR spectrum components; (c) visible component time profiles; (d) NIR component time profiles. Extracted decay time constants were determined using a mono-exponential fitting (orange dashed line).}
  \label{fig:pcfsTAS data}
\end{figure*}

The transient excited state spectrum of Pc:PTP exhibits many similar features to that of DAP:PTP, with a negative GSB at 508 and 546 nm, a broad ESA superimposed throughout the observable spectrum and additional negative bands at 600 nm and 644 nm matching the steady-state fluorescence spectrum (Figure~\ref{fig:CombinedOpticalProperties_Final}(d)). A clear difference compared to DAP:PTP is that the decay of these features extends beyond the 5.9 ns termination of the measurement. Furthermore, our measurements and subsequent analysis by SVD were unable to discern any bands that could sensibly be associated with the triplet state other than the in-growing GSB (Extended Figure~\ref{fig:pcfsTAS data}). In this instance, while two components were required to sufficiently account for the visible spectrum, only one component was found to harbour a significant weighting for a description of the NIR region. All components return mono-exponential decay rates between 4.5 and 10 ns. Here, the first component of the visible region represents the only in-growing component. Therefore, similar to DAP:PTP, this likely represents a growing abundance of $T_{1}$ states and the associated 7 ns lifetime can be approximated as $\tau_{ISC}$ (giving $\kappa_{ISC}$ = 0.143 ns$^{-1}$). However, the observed spectrum only represents a partial relaxation of $S_{1}$ states, so this is likely an underestimation. In the literature, $\tau_{f}$ for room-temperature Pc:PTP is given as a value between 8 - 10 ns, which is closely matched by the 10 ns lifetime of the second visible component\cite{Patterson1984,Kryschi1992,Williams1983}. Therefore, we expect $\tau_{ISC}$ to be longer than 8 ns\cite{Brouwer1999}. The search for triplet state adsorptions for Pc:PTP would then require nsTAS measurements, which were beyond the scope of this work. We note that within the frame time of the experiment, the spectra of both Pc:PTP and DAP:PTP were devoid of triplet absorptions at 520 and 510 nm, respectively, which were previously used by Kouno \textit{et al.,} 2019 to determine the triplet lifetimes\cite{kouno2019}. Together, these data comprehensively demonstrate that the triplet quantum yield is markedly higher for DAP:PTP than Pc:PTP, which is a favourable attribute for achieving a higher amount of spins and hence a higher gain in maser devices.

\subsection{Density Functional Theory Calculations}

To understand how the chemical structures of Pc and DAP might relate to their observed optical properties, we conducted time-dependent density functional theory (TD-DFT) quantum calculations to observe the singlet and triplet states. Prior to excited state calculations, each molecule was optimised to find their respective singlet and triplet geometries using a polarisable continuum model (PCM) of benzene to approximate the non-polar environment of para-terphenyl (PTP)\cite{Bogatko2016}. To provide a more intuitive depiction of the states relative to each other, the calculated energies were normalised against the energy of the ground state, such that the energy of $S_{0}$ = 0 eV. The energy of each successive state is then given by their calculated excitation energy. Using this approach, the experimental trends were reproduced with the $S_{0}$ to $S_{1}$ excitation energy of DAP red-shifted by 30 nm (0.2 eV) compared to Pc (see the expanded Jablonski diagram in Figure~\ref{fig:DFT-derived State Energies}(a)), while the absolute excitation energies are underestimated by approximately 15\%, which is typical of TD-DFT calculations. The energy gap between $S_{1}$ and $T_{2}$ ($\Delta E_{ST}$) can be estimated as 0.12 eV for DAP, compared with 0.27 eV for Pc. Since unfunctionalised linear aromatic hydrocarbons only exhibit weak spin-orbit coupling\cite{Pedash2002, Schott2017}, $\Delta E_{ST}$ is closely linked to $\kappa_{ISC}$ following Fermi's golden rule\cite{Marian2021}. Thus, even small differences in $\Delta E_{ST}$ can have profound impacts on $\kappa_{ISC}$ and $\phi_{T}$. For example, $\kappa_{ISC}$ has been shown to differ by two orders of magnitude as a result of a 20 meV difference in $\Delta E_{ST}$ between alternative Pc lattice sites in a PTP crystal host\cite{Patterson1984}. While TD-DFT-based estimations of excited state energies are useful for understanding the relative $\Delta E_{ST}$ of different Pc derivatives, \textit{ab initio} calculations of DAP and Pc comparing the use of a PCM to an explicit host model have demonstrated that the latter is required to properly account for the influence of the host\cite{Charlton2018}. Hence, due to additional weak charge transfer interactions between DAP and PTP described by Bertoni \textit{et al.,} our estimated $\Delta E_{ST}$ likely represents a small overestimation \cite{Bertoni2022}.

Since spin-orbit coupling in organic molecules is mainly derived from transitions involving non-bonding orbitals, we considered if the 6,13-position nitrogen atoms may influence $\kappa_{ISC}$ through incorporation of their lone pair p-orbitals. TD-DFT calculations of DAP and Pc reveal that the transition between $S_{0}$ and $S_{1}$ only involves the HOMO and LUMO. An analysis of the natural transition orbitals for DAP indicate very similar $\pi$-$\pi^*$ character devoid of out-of-plane non-bonding contributions, similar to Pc. Hence, according to the El-Sayed approximation where $\kappa_{ISC}$ depends on a change in orbital type\cite{ElSayed63,Marian2021}, the nitrogen substitution at the 6,13-positions is not expected to contribute significantly to the formation of triplets. Indeed, further calculations indicate that DAP exhibits only a weak degree of spin-orbit coupling between singlet and triplet states, particularly between $S_{1}$ and $T_{3}$ states (Refer to Table S I in Supplementary Information). Overall then, it is expected that the increased $\kappa_{ISC}$ observed for DAP relative to Pc is principally due to the reduced $\Delta E_{ST}$.
\subsection{Zero-field Transient EPR}

\begin{figure}[h]
  \includegraphics{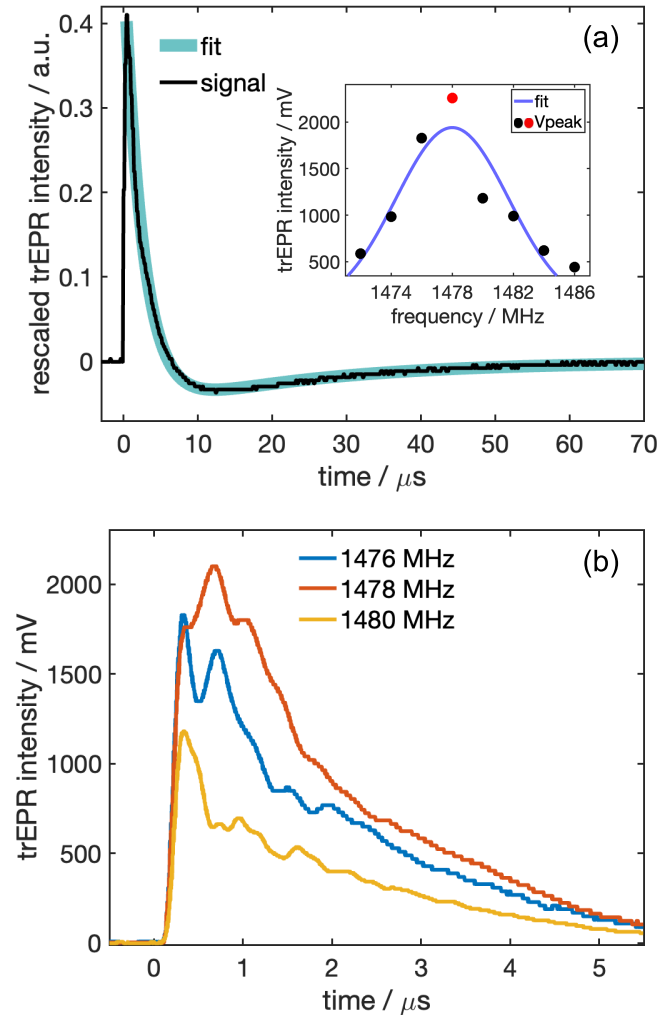}
  \caption{(a) The ZF-trEPR signal of DAP:PTP measured at 1478 MHz (black curve). This signal is rescaled with the peak voltage equal to the maximum population difference between $T_z$ and $T_x$ (0.41\cite{Bogatko2016}). A biexponential fitting was plotted (green curve) in order to estimate the spin dynamics. Other measured peak voltages are plotted in the inset figure and fitted with a Gaussian curve. (b) The signals measured at 1476 MHz, 1478 MHz, and 1480 MHz, with the signal at 1478 MHz being the highest.}
  \label{fig:trEPR}
\end{figure}

Before attempting to mase DAP:PTP on its $T_x-T_z$ transition, the transition frequency was found using ZF-trEPR when the sample was excited by a 620-nm laser pulse. The strongest ZF-trEPR signal (taken to be the resonant signal) was found at 1478 MHz (Figure~\ref{fig:trEPR}(a)), where the signal was initially emissive for 7 $\mu$s before shifting to being absorptive for another 30 $\mu$s. The shape of the resonant signal at a shorter time span is shown in Figure~\ref{fig:trEPR}(b), while comparing it to two signals at adjacent frequencies. Since the experiments were conducted with a 2 MHz step size, the frequency of $T_x-T_z$ transition was determined to be $f_{T_x-T_z}=1478 \pm 2$ MHz. Some oscillations can be seen for the resonant signal in Figure~\ref{fig:trEPR}(b) between 0 and 1 $\mu$s. Since they persist even at low microwave powers and do not change frequency with microwave power (see Extended Data Figure~\ref{fig:power}), the oscillations are likely due to noise.
\begin{figure*}[htbp]
  \includegraphics[width=\textwidth]{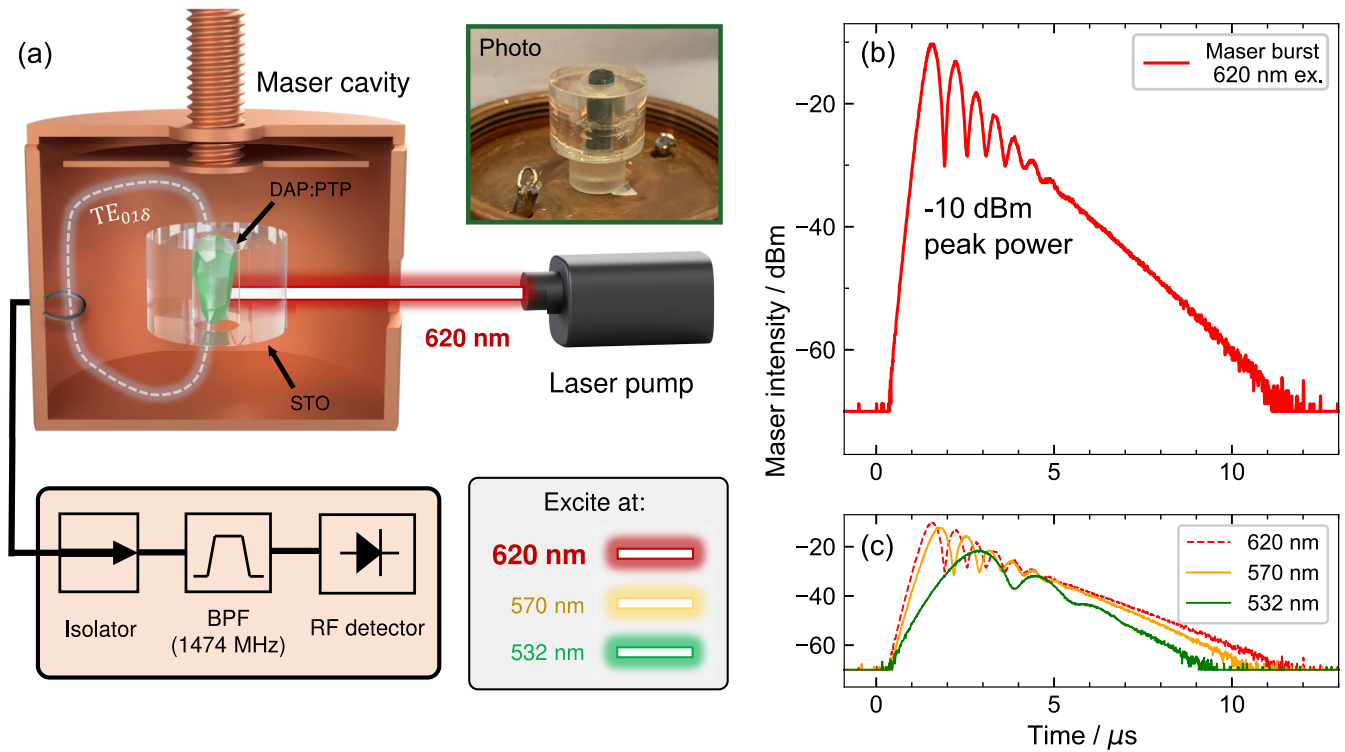}
  \centering
  \caption{(a) Cross-section rendering of the maser cavity containing the DAP:PTP crystal, dielectric STO resonator and surrounding copper cavity. The STO resonator supported a TE$_{01\delta}$ mode at a frequency $f_{\text{mode}}$ = 1478 MHz. The DAP:PTP crystal was side-pumped through a hole in the copper cavity with 620, 570 and 532 nm laser light. The maser output was then coupled through a copper loop into an isolator, bandpass filter (BPF) and RF detector (a log detector) which could be read on an oscilloscope. A photo of the DAP:PTP crystal in the STO resonator is shown. (b) The output maser signal of DAP:PTP obtained with 620-nm optical excitation. The peak output is -10 dBm, and the Rabi oscillations have a frequency of 1.6 MHz. (c) Comparison between DAP:PTP masing signals (in dBm) using different excitation wavelengths.}
  \label{fig:maseroutput}
\end{figure*}
The triplet spin dynamics can be elucidated by simulating the observed ZF-trEPR signals, which are related to the populations of the triplet sublevels (as sketched in Figure~\ref{fig:simplejablon}(a)). The dynamics of the third sublevel, $T_y$, is assumed to be negligible. The voltages of the ZF-trEPR signals are then proportional to the population difference between $T_x$ and $T_z$, denoted as $N_x(t)-N_z(t)$. The time evolution of $N_x$ and $N_z$ can be represented by a group of differential equations:
\begin{equation}
    \begin{bmatrix}
\dot{N}_x(t)\\\dot{N}_z(t)
\end{bmatrix} =
\begin{bmatrix}
-w_{xz}-k_x& w_{xz}\\
w_{xz}& -w_{xz}-k_z
\end{bmatrix} 
\begin{bmatrix}
{N}_x(t)\\{N}_z(t)
\end{bmatrix}
\label{Eqn:dynamics}
\end{equation}
where $w_{xz}$ is the spin-lattice relaxation rate between $T_x$ and $T_z$, $k_x$ and $k_z$ are the depopulation rates of these two sublevels \cite{Matsuoka_biexp}. The upward and downward spin-lattice relaxation rates are assumed to be the same at room-temperature since they are related by the Boltzmann factor\cite{Hintze2017}.

The signal at 1478 MHz was rescaled such that the maximum amplitude was 0.41 (Figure~\ref{fig:trEPR}(a)), corresponding to the maximum population difference (i.e. $N_x(0)-N_z(0)=0.41$ with $N_x(0)+N_y(0)+N_z(0)=1$) between those two sublevels \cite{Bogatko2016}. A biexponential fitting was then applied to the signal as in Equation~\ref{eqn:biexp}, in order to find the relaxation rates: 
\begin{align}
    N_x(t)-N_z(t)&=A\exp(\alpha_-t)+B\exp(\alpha_+t)
    \label{eqn:biexp}\\
    \alpha_{\pm}&=-(w_{xz}+k_{avg})\pm\sqrt{w_{xz}^2+(\Delta k)^2}
    \label{Eqn:eigenvalues}
\end{align}
where $\alpha_-$ and $\alpha_+$ are two eigenvalues of the coefficient matrix in Equation~\ref{Eqn:dynamics}. $k_{avg}=(k_x+k_z)/2$ and $\Delta k=(k_x-k_z)/2$. The simulated curve is shown as the green curve in Figure~\ref{fig:trEPR}(a), with the fitted parameters being $A=0.547\pm0.003$, $B=-0.066\pm0.005$, $\alpha_-=(-3.93\pm0.06)\times10^{5}s^{-1}$ and $\alpha_+=(-0.459\pm0.017)\times10^{5}s^{-1}$.

According to Equation~\ref{Eqn:eigenvalues}, the combination of two depopulation rates can be calculated as
\begin{equation}
    w_{xz}+\frac{k_x+k_z}2=-{\frac{1}{2}}(\alpha_-+\alpha_+)
\end{equation}
which corresponds to a decay rate of $ 0.2\times10^6\ s^{-1}$ (decay time 4.6 $\mu$s). This gives an upper limit on the decay time of the spins, as it is not possible to separate the spin-lattice relaxation rate and depopulating rates from this simulation without knowing them independently.

\subsection{Maser and cQED experiments}
\begin{figure*}[htbp]
  \includegraphics[width=\textwidth]{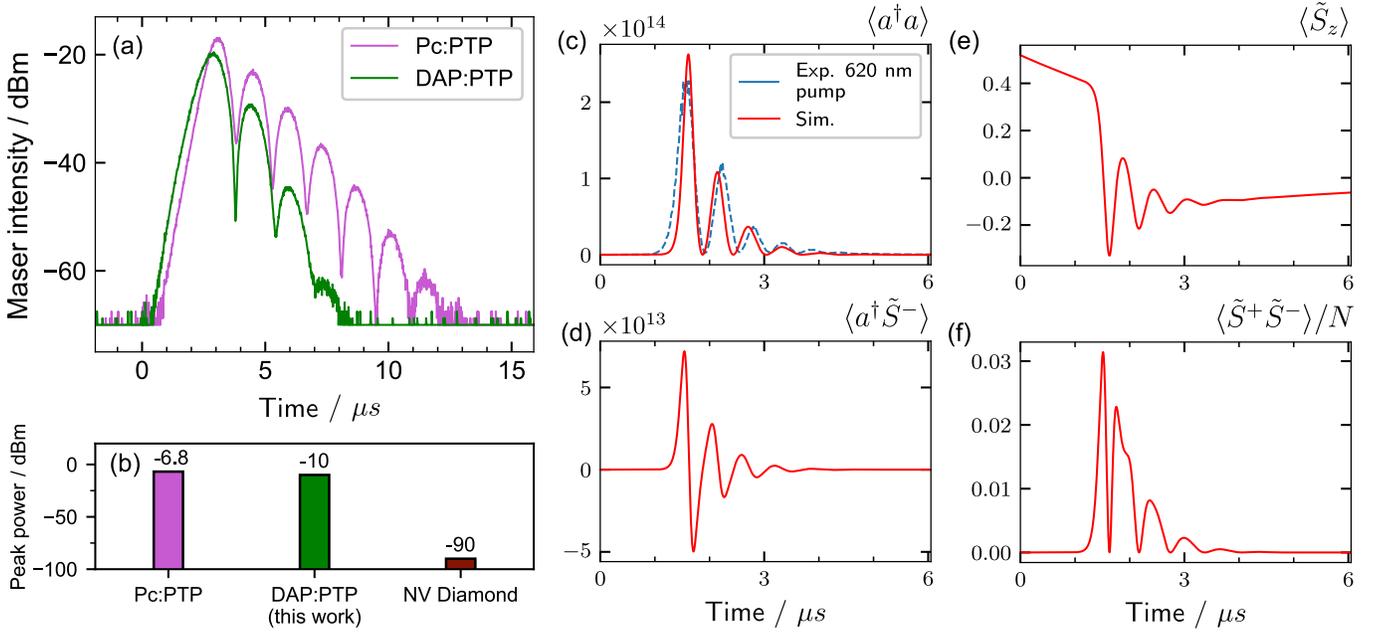}
  \centering
  \caption{(a) Comparison of maser bursts from DAP:PTP and Pc:PTP when excited under identical conditions and optical pumping. Only the cavity frequency was adjusted to match their respective resonances. (b) Comparison of highest reported maser powers for the known room-temperature masers. Simulated expectation values of the (c) cavity photon number $\langle a^\dagger a\rangle$, (d) the complex values of the spin-photon coherence $\langle a^\dagger \Tilde{S}^-\rangle$, inversion $\langle \Tilde{S}^z\rangle$ and spin-spin correlation $\langle \Tilde{S}^+\Tilde{S}^-\rangle/N$ during the maser burst from DAP:PTP in Figure~\ref{fig:maseroutput}(b) are also shown here.}
  \label{fig:dappc}
\end{figure*}
In order to reduce the laser pumping threshold of DAP:PTP required to mase, a microwave cavity with a high loaded quality factor ($Q_L$) and low mode volume ($V_{\text{mode}}$) was constructed in order to boost the spin-photon coupling strength via the Purcell effect\cite{breeze2015enhanced}. Figure~\ref{fig:maseroutput}(a) shows the cavity consisting of an outer copper cavity enclosing the transparent strontium titanate (STO) dielectric ring resonator. The TE$_{01\delta}$ mode of the STO resonator resonated at $f_{\text{mode}}=1478$ MHz, which could be tuned by raising/lowering the ceiling of the copper cavity via turning a screw.

The DAP:PTP single crystal was then placed in the STO ring and pumped with a 5 ns laser pulse at 620 nm, with a repetition rate of 10 Hz and average energy of 13.5 mJ per pulse. Using a coupling wire loop attached to a coaxial cable, the maser output was coupled out from the cavity and passed through an isolator and bandpass filter before detection. The power envelope of a single shot of the maser burst is shown in Figure~\ref{fig:maseroutput}(b), lasting for approximately 10 $\mu$s (laser trigger at 0 $\mu$s). The cavity was tuned to a frequency $f_{\text{mode}}$ that maximised the maser output, which was found to be 1474$\pm$1 MHz. The frequency was measured using a vector network analyser (VNA) to read the $S_{11}$ reflection dip of the cavity at the frequency of maximum maser output, with an estimated error of 1 MHz due to the frequency span of the VNA limiting precision. The difference between the masing signal frequency and the ZF-trEPR frequency ($1478\pm2$ MHz) is likely due to slight heating of DAP:PTP from constant laser excitation in an enclosed insulating cavity, which lowered the resonant frequency. The maser burst features Rabi oscillations due to the strong coupling between the DAP:PTP spins and the microwave photons of the electromagnetic mode, and have a measured Rabi frequency of 1.6 MHz.

Figure~\ref{fig:maseroutput}(c) shows the comparison between maser bursts using different pump wavelengths (with identical pulse energy and repetition rates) at 570 nm (corresponding to an absorption peak in Figure~\ref{fig:CombinedOpticalProperties_Final}(a)) and 532 nm, the wavelength of frequency-doubled Nd:YAG lasers. Since 620 nm was the wavelength of highest peak absorption in Figure~\ref{fig:CombinedOpticalProperties_Final}(a) and lower wavelengths featured lower absorption peaks, the maser power output correspondingly decreases at shorter wavelengths for DAP:PTP due to lower absorption of the pump power (i.e. fewer triplet spins generated). However, all pump wavelengths still give substantial maser powers at -12 dBm for 570 nm excitation and -21 dBm for 532 nm excitation. The maser bursts also reach peak powers at later times for shorter wavelengths of laser excitation; 1.5 $\mu$s for 620 nm excitation and 3 $\mu$s for 532 nm excitation.

Figure~\ref{fig:dappc}(a) compares the maser outputs between DAP:PTP and Pc:PTP. The cavity setup, $Q_L$ and measuring apparatus were identical as in Figure~\ref{fig:maseroutput}(a), except for how both crystals were excited with 532 nm pulsed light with an average energy of 7 mJ per pulse, and the cavity was tuned near 1450 MHz for the Pc:PTP maser output. The trigger time of the laser in Figure~\ref{fig:dappc}(a) is the same between the two materials, hence DAP:PTP is seen to reach its peak maser power faster than Pc:PTP. The maser output of Pc:PTP however lasts slightly longer than that of DAP:PTP; 12 $\mu$s for the former compared to 8 $\mu$s for the latter. This may be due to the faster spin-lattice relaxation and/or depopulating rates in DAP:PTP, which is evident from its shorter ZF-trEPR signal in Figure~\ref{fig:trEPR} compared to that of Pc:PTP, which lasts for about 40 $\mu$s\cite{HaoWu2019}. Finally, both preserve Rabi oscillations for the duration of the maser burst.

When not constrained to using the same pumping wavelength, Figure~\ref{fig:dappc}(b) shows the highest peak maser powers of DAP:PTP and the only two other known room-temperature masers when pumped with their ideal wavelengths. Pc:PTP still has the highest reported peak power (-6.8 dBm\cite{Breeze2017cqed}) when pumped with 592 nm light, but DAP:PTP is quite close at -10 dBm when pumped with 620 nm light. Importantly, DAP:PTP can achieve this peak power with less than half the loaded $Q$ factor used in the Pc:PTP masing experiment (3690 for the former versus 8500 for the latter) and slightly less pump energy (13.5 mJ for DAP:PTP versus 15 mJ for Pc:PTP)\cite{Breeze2017cqed}, which may denote a lower masing threshold for DAP:PTP when pumped with 620 nm light. Both Pc:PTP and DAP:PTP feature powers much higher than that reported for NV$^-$ diamond, whose highest reported power was -90 dBm\cite{Breeze2018} under continuous operation. Though the peak power of a diamond maser under pulsed operation may be higher than -90 dBm, this has yet to be reported. Finally, it is noted that the masing bursts for DAP:PTP and Pc:PTP are able to follow the repetition rate of the pump laser, and hence the masing signals seen in Figure~\ref{fig:maseroutput} and \ref{fig:dappc} have a repetition rate of 10 Hz which follows the pumping repetition rate.

We now explore whether the masing of DAP:PTP in the microwave cavity has managed to reach the strong coupling regime vital for cQED experiments. The strong coupling regime can be reached when the ensemble spin-photon coupling ($g_e$) of a system greatly exceeds the cavity decay ($\kappa_c$) and spin-dephasing ($\kappa_s$) rates. This can be summarised by the `cooperativity' of the system defined as $C=4g_e^2/(\kappa_c\kappa_s)$\cite{Breeze2017cqed}, where strong coupling is reached if $C\gg1$. $\kappa_c$ can be calculated to be $2\pi f_{mode}/Q_L\approx2.51$~MHz from the cavity setup, while the other two parameters $\kappa_s$ and $g_e$ can be estimated by simulating the maser burst in Figure~\ref{fig:maseroutput}(b). This is done using a set of coupled differential equations derived from the master equation with a Tavis-Cummings Hamiltonian and the Liouvilian components accounting for the various decay pathways. These differential equations are the same as in previous literature\cite{Breeze2017cqed} for simulating the expectation values of the cavity photon number, spin-photon coherence, inversion and spin-spin correlation ($\langle a^\dagger a\rangle$, $\langle a^\dagger \Tilde{S}^-\rangle$, $\langle \Tilde{S}^z\rangle$ and $\langle \Tilde{S}^+\Tilde{S}^-\rangle/N$ respectively), but with different initial parameters (see Supplementary Information).

The maser burst can be converted to photon numbers through the following equation:
\begin{equation}
    \langle a^\dagger a\rangle=P(t)\frac{1+K}{hf_{mode}\kappa_cK}
    \label{eqn:adaggera}
\end{equation}
where $P(t)$ is the maser output power in units of watts and $K$ is the cavity coupling coefficient (measured to be 0.20 using VNA methods\cite{kajfez1995,kajfezwebsite}, see Supplementary Information). The coupled differential equations are solved so that they fit to the observed $\langle a^\dagger a\rangle$ calculated from Figure~\ref{fig:maseroutput}(b), in order to generate the simulated expectation values in Figure~\ref{fig:dappc}(c)-(f). The resulting parameters $\kappa_{s,sim}$ and $g_{e,sim}$ (denoted as such due to being simulated) were $2\pi\times$0.29 MHz and $2\pi\times$2.3 MHz respectively. This gives a cooperativity of $C=182$, which is about two orders of magnitude higher than that experimentally measured with collective NV$^-$ spins\cite{Eisenach2021}, revealing the superiority of molecularly-doped spin systems in terms of achieving strongly coupled spin-photon systems at ambient conditions. 

We note however that there is a discrepancy between the simulated $g_{e,sim}$ and its relation to the observed Rabi frequency ($\Omega$) of the maser burst. In previous studies of strongly coupled spin-photon systems, the ensemble spin-photon coupling strength $g_e$ is equal to $\Omega/2$\cite{Amsuss2011,Kubo2011}. Based on this relationship and $g_{e,sim}=2\pi\times2.3$ MHz, the Rabi frequency arising from the strong coupling between the spins and microwave photons in our experiment can be predicted to be $2\pi\times4.6$ MHz. However, the Rabi frequency observed in our experiments was $\Omega=2\pi\times1.6$ MHz, which is almost three times slower than the predicted value. We attribute the discrepancy to the large number of excitations (i.e. microwave photons) in the cavity, $n_{cavity}$, arising from the masing process. It has been found that when $n_{cavity}\gg1$ and comparable to the number of spins $N$, the Rabi frequency will be proportional to $\sqrt{n_{cavity}}$\cite{Eisenach2021} which indicates the transition of the strong coupled spin-photon system from the quantum to the classical regime\cite{chiorescu2010magnetic}, where the photon can be treated as a classical field normally adopted in magnetic-resonance studies. 

\section{Discussion}

We have characterised the optical and zero-field spin dynamics of DAP:PTP, where it was shown to have an advantageously higher triplet quantum yield and faster $\kappa_{ISC}$ than Pc:PTP, alongside being able to absorb longer wavelengths of pump light. 

When operating as a maser and as a platform for room-temperature cQED, DAP:PTP was able to mase using three different pumping wavelengths; 532 nm, 570 nm and 620 nm, the lattermost being its ideal pumping wavelength. Through this, we suspect that it can mase using any wavelength between 532 and 620 nm. When analysed using the master equations, DAP:PTP gave a cooperativity of 182, which places it in the strong coupling regime for cQED at room-temperature. We found that DAP:PTP has compelling advantages against Pc:PTP; firstly, 620 nm light, which is most suitable to pump DAP:PTP, is easier to generate efficiently than shorter wavelength 590 nm light that is ideally used to pump Pc:PTP. Secondly, DAP:PTP has a faster masing startup time (less lag), and possibly has a lower masing threshold which makes it easier to mase using less pumping power or lower $Q_L$. We hypothesise that this lower threshold may be related to the higher triplet quantum yield and faster ISC rate it possesses, which would translate to more spins per optical pulse being able to participate in masing. Thirdly, DAP is also more chemically stable than Pc, since the presence of nitrogen groups makes it less air-sensitive. This would make doping it into PTP easier since it would not decompose as easily during crystal growth processes. 

When compared head-to-head using pulsed 532 nm excitation (which is also the first time Pc:PTP has been reported to mase under 532 nm light), both materials perform almost equally well. DAP:PTP's sole disadvantage would be its shorter spin-lattice relaxation time (shown by ZF-trEPR experiments) which causes its maser burst to decay faster than Pc:PTP under an identical excitation time, but it still maintains its faster startup time to reach peak amplitude. We believe the advantage of a faster startup time will outweigh the disadvantageous relaxation time since DAP:PTP could be used for microwave amplification of even faster signals than Pc:PTP, whereas the time of the maser signal could be lengthened by just using a longer optical pump pulse. Indeed, attaching a luminescent concentrator to DAP:PTP to attempt continuous-wave masing using longer pump pulses, as was performed with Pc:PTP\cite{HaoWu2020_cont}, is a subject of future work, where its lower masing threshold may allow it to perform better.

In conclusion, DAP:PTP is a promising new room-temperature maser gain medium which can be excited with red (620 nm) light as well as green (532 nm) light, the latter being a common wavelength of workhorse frequency-doubled Nd:YAG pulsed lasers. As only the second material discovered to be able to mase at both zero-field and room-temperature in almost a decade, DAP:PTP can help further the search for other maser materials that share its properties, while acting as an attractive platform for studying room-temperature cQED with high cooperativity.

\section{Methods}
\label{Sec:Methods}

\subsection{Synthesis of DAP and DAP:PTP crystal growth}
DAP was synthesised by co-melting 2,3-diaminonaphthalene (AlfaAesar, 97\%) with naphthalene-2,3-diol (Fluorochem, 95\%) at 216 $\degree$C  under argon for 1 hour. The mixture was then cooled and formed into a slurry by vigorous stirring with acetone for 30 minutes. This solution was filtered to yield crude 6,13-dihydro-6,13-diazapentacene as brown powder which was heated in toluene under argon with chloranil (SigmaAldrich, $\geq$98\%)  for a further hour. The solution was filtered hot and washed with excess aqueous Na$_{2}$CO$_{3}$ to yield DAP as a dark green powder (80\%). Prior to mixing with PTP, DAP was purified by vacuum sublimation at 220 $\degree$C. Dark green crystals formed after 2 days.   

\begin{figure}[htbp]
    \centering
    \includegraphics[width=\linewidth]{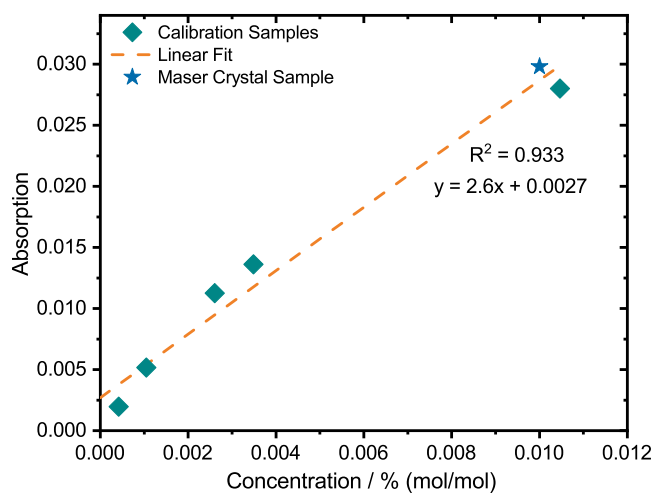}
    \caption{[Extended Data Figure] UV/Vis calibration curve used to estimate the concentration of the DAP:PTP maser crystal. Standard solutions were made using an ortho-terphenyl host. The concentration of the crystal used for maser experiments was estimated by diluting a representative piece of the DAP:PTP crystal with ortho-terphenyl until the sample adopted a transparent appearance as with the calibration standards.}
    \label{fig:calibcurve}
\end{figure}

Samples of Pc:PTP and DAP:PTP were grown using the Bridgman method described previously\cite{Sloop1981, Ai2017}. PTP (Sigma-Aldrich, $\geq$99.5\%) was extensively purified by zone refining prior to use. DAP was mixed with PTP at 0.05\% concentration by grinding in a pestle and mortar before being loaded into a 3.5 mm OD borosilicate NMR tube and sealed under argon. Bridgman growth was performed at 4 mm/hr over 3 days at 218 $\degree$C. Unlike Pc:PTP which grows as a homogeneous pink crystal, DAP:PTP grew as a banded crystal with a relatively clear section at the bottom, a homogeneous green section in the middle and a smaller concentrated section at the top. For maser experiments, the top and bottom bands were discarded and the concentration of the middle section was determined using UV/Vis spectroscopy on samples of DAP in a clear glass of ortho-terphenyl at known concentrations. A calibration curve was constructed using absorbance at 614 nm, and the concentration was subsequently determined to be ca. 0.01\% (Extended Figure~\ref{fig:calibcurve}). All maser and ZF-trEPR experiments used DAP:PTP and/or Pc:PTP with 0.01\% dopant concentration.

\subsection{TCSPC}
The time-resolved photoluminescence (PL) spectroscopy was measured with TCSPC which gave carrier recombination lifetimes of materials. A diode laser (NanoLED, HORIBA) was used to produce 1 mW cm$^{-2}$, $<$200 ps excitation pulses at a repetition rate of 1 MHz. 404 nm and 635 nm lasers were used as the excitation sources. A photomultiplier tube detector (PPD-900) collected photons at the desired wavelengths. The instrument response function was measured with a pristine sample substrate at the excitation wavelength. Additional long-pass filters were used before the detector to exclude scattering from the excitation source for some measurements.

\subsection{Femtosecond Transient Absorption Spectroscopy}
Measurements were done using a broadband pump-probe commercial spectrometer (Helios, Spectra Physics, Newport Corp). Ultrafast laser pulses (800 nm, $<$100 fs pulse duration) were generated by a 1 kHz Ti:sapphire regenerative amplifier (Solstice, Spectra Physics, Newport Corp). The excitation pump and white light probe were converted from 800 nm pulses. The wavelength tuneable pump light (400 nm in this work) was converted by an optical parametric amplifier (TOPAS Prime, Spectra-Physics) and a frequency mixer (Niruvis, Light Conversion). The white light probe was generated by another part of 800 nm pulses directed through a crystal (sapphire for the visible region and yttrium aluminium garnet for the NIR region). To reduce the noise, the white light was split into two beams, where one is passed through the sample, and the other is used as a reference. 

Pump and probe light was focused on the sample with a non-collinear optical path and overlapped at the sample position. The time delay between the pump and probe is controlled by a mechanical delay stage with a time range of 6 ns. The pump light is modulated with 500 Hz chopper. The time-resolved transmission spectrum changes between pump on and pump off are measured by fibre-optic coupled multichannel spectrometers (CMOS sensors). The beam sizes of the focused probe and pump pulses are 0.5 mm$^2$ at the sample position. The pump light energies were measured with an energy meter (VEGA P/N 7Z01560, OPHIR Photonics). To reduce probe noise from laser fluctuation, the measured transmission spectrum difference was normalised to the reference probe spectrum and averaged for several scans to achieve a good signal-to-noise ratio. SVD analysis was performed using SurfaceExplorer (v4.3.0). Spectral and principal component data were then analysed in OriginPro 2022b. Decay values were found by fitting a mono-exponential decay function.

\subsection{TD-DFT calculations}
Electronic calculations including geometry optimisations of the ground singlet and triplet states were performed using Gaussian16 software at the B3LYP 6311G''dp level. Natural transition orbitals were generated by specifying [pop=(nto, savento) density=transition=1]. To check the reliability of our data for comparing DAP and Pc, their relative shift was compared to the UV/Vis spectrum, where a similar 30 nm shift in $S_{0}$ to $S_{1}$ is also observed. Additional calculations to estimate the singlet to triplet spin-orbit coupling were performed in Orca 5.0.1 software using previously optimised singlet state structures. Here, a TD-DFT B3LYP def2-TZVP basis set was employed.

\subsection{ZF-trEPR setup}

A homemade trEPR spectrometer was constructed with a configuration similar to that in previous work \cite{HaoWu2019} which utilised an LC-tank resonator and I-Q homodyne detection to detect EPR signals from the sample. An optical parametric oscillator (OPO, Litron Aurora II Integra) pumped by its own internal Q-switched Nd/YAG laser was used to excite a small 0.01\% DAP:PTP crystal (3 x 1 x 1 mm dimensions) at 620 nm with pulse length 10 ns and repetition rate 10 Hz. The light was passed through a convex lens and converged on the sample which was placed at the center of the handmade inductor coil resonator. Through using optical filters, the pumping energy was adjusted to approximately 0.6 mJ per pulse, measured with a laser power meter (Thorlabs PM100D). A variable capacitor (BB833) controlled by a DC bias was used to adjust the resonant frequency. 

\begin{figure}[htbp]
    \centering
    \includegraphics[width=\linewidth]{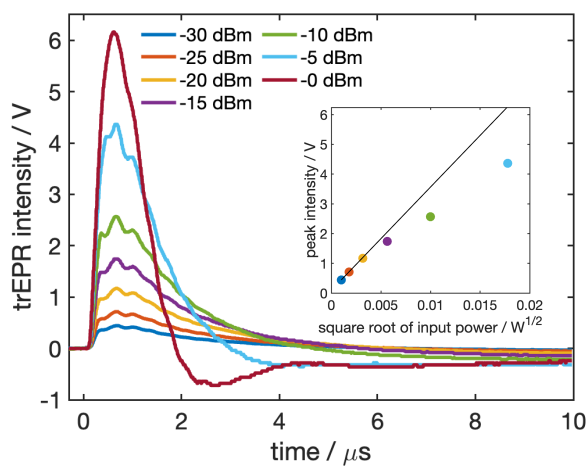}
    \caption{[Extended Data Figure] ZF-trEPR signals of DAP:PTP collected under the same experimental conditions as in Figure~\ref{fig:trEPR}, but with different input microwave powers. Oscillations occur at all microwave powers. In the inset plot, the relation between peak trEPR voltage and square root of the input microwave power (in watts) is illustrated. The proportional relation is violated when the input power is higher than -20 dBm, which indicates the onset of power saturation.}
    \label{fig:power}
\end{figure}

The signals were captured on an oscilloscope (Rigol DS1104 Z-plus) triggered by a photodiode (Thorlabs DET210). Each signal plot in Figure~\ref{fig:trEPR} is an average of 512 signals collected on the oscilloscope at a specific microwave frequency. In Extended Figure~\ref{fig:power}, the microwave power was varied to study the relation between the EPR signal voltage and the input microwave power in order to avoid saturation. Once a suitably low enough microwave power was found to avoid saturation, which was -25 dBm (0.00316 mW), this power level was used for the experiments in Figure~\ref{fig:trEPR}.

\subsection{Maser setup}

\begin{figure*}[h]
    \centering
    \includegraphics[width=12cm]{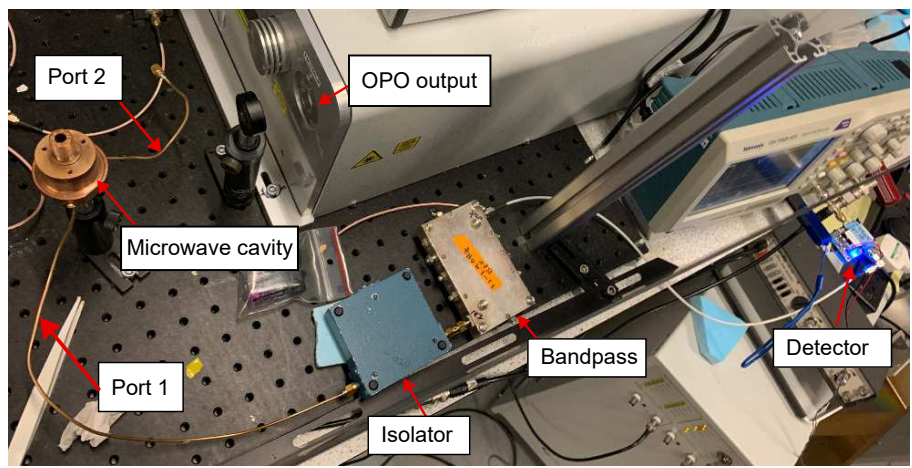}
    \caption{[Extended Data Figure] Photo of the maser setup; the laser pulses from the OPO go through a lens and into the microwave cavity to excite the DAP:PTP crystal within (an image of the DAP:PTP and dielectric resonator inside the copper cavity is shown in Figure~\ref{fig:maseroutput}). The maser burst can then be read from Port 1 after passing through an isolator and bandpass.}
    \label{fig:rig}
\end{figure*}

The setup, pictured in Extended Figure~\ref{fig:rig}, consists of the 0.01\% DAP:PTP single crystal (cylinder-shaped, 4 mm diameter and 8 mm height) inserted into a microwave cavity, which was connected to a microwave isolator and bandpass filter before being connected to a logarithmic amplifier which acts as a detector.

The cavity consisted of a dielectric ring of a single crystal of strontium titanate (STO) housed within a cylindrical copper cavity. The STO ring (OD 12 mm, ID 4.05 mm and height 8.6 mm) was placed on a cross-linked polystyrene stand that raised it to 3 mm above the floor of the copper cavity. The copper cavity had an inner diameter of 30 mm and maximum inner height of 20.4 mm. An inductive loop, made by soldering a wire loop onto the end of an SMA cable (see photo in Figure~\ref{fig:maseroutput}), was inserted into the cavity to act as an output coupler for the maser burst. This coupling loop (called Port 1 in Extended Figure~\ref{fig:rig}) had a coupling coefficient of $K=0.20$ (undercoupled, refer to Supplementary Information), and would have been coupled to the TE$_{01\delta}$ mode of the STO ring. The frequency of this mode could be adjusted by using a tuning screw to raise/lower the inner ceiling of the copper cavity. A second coupling loop (similar in make to the first and denoted as Port 2 in Extended Figure~\ref{fig:rig}) was also inserted into the cavity for monitoring the frequency using a VNA (HP8753A, lightly calibrated before measurements). This second coupling loop was severely undercoupled ($K\ll0.20$) to prevent reducing $Q_L$. When measuring and recording the masing signal, this second coupling port was left disconnected from everything. 

Besides the main circular hole (5 mm diameter) used for allowing light to enter the cavity and pump the crystal from the side, the copper cavity featured slot gaps which were cut into its main body. These slot gaps (which were unrelated to masing experiments) were actually causing radiative loss for the microwave cavity which reduces $Q_L$, which is detrimental to masing. If one desires to make a copper cavity for masing, it is best to avoid having so many holes in the resonator walls and reduce radiative losses (holes for pump light entry are acceptable, but not too large). However, we note that, even with this unoptimised copper cavity, a maser signal can still be seen easily.

The isolator mitigates reflections of the maser pulse, but we note that a maser pulse can still be observed with the isolator removed (in Extended Figure~\ref{fig:rig}, the maser burst can be immediately detected from the end of Port 1). The bandpass filter, with a pass band between 1310 and 1880 MHz, was used to narrow down the frequency detection range since the logarithmic amplifier detection range was very wideband. The overall attenuation of the isolator (forward direction) and bandpass filter with coaxial cables was negligible, amounting to less than 1 dB. 

Optical pumping of DAP:PTP for masing was supplied using the same OPO used for ZF-trEPR with the same pulse width and repetition rate. For Figure~\ref{fig:maseroutput}(b) and (c), the pulse energy (for all wavelengths) was adjusted such that each pulse was 13.5 mJ on average, with a standard deviation of 0.7 mJ. In Figure~\ref{fig:dappc}(a), the pulse energy was 7 mJ on average, with a standard deviation of 1 mJ. The laser spot size at the crystal was 2 mm in all maser experiments.

The experiment proceeded as follows; the DAP:PTP crystal (or Pc:PTP crystal) was pumped continuously using the OPO for a short time to let the cavity resonant frequency stabilise (the resonance drifts due to heat from lasing). After using the VNA to check that the frequency had mostly stabilised, the VNA was disconnected and the cavity ceiling height was tuned until the largest amplitude maser pulse with the highest frequency of Rabi oscillation was found. The VNA was then reconnected to quickly record down the cavity frequency that gave this ideal maser pulse, after which it was again disconnected and a single shot of the maser pulse was read from the detector, as displayed in Figures~\ref{fig:maseroutput} and \ref{fig:dappc}. The logarithmic amplifier can't reliably detect signals below -70 dBm, and so the maser plots are limited by a -70 dBm lower bound.
\subsection{cQED simulations}
The maser signal in Figure~\ref{fig:maseroutput}(b) was converted into watts, and by using Equation~\ref{eqn:adaggera} and the calculated $\kappa_c$ and $K$ values, the experimental dataset of $\langle a^\dagger a\rangle$ could be calculated. The dataset was baseline corrected so that its initial values before masing were set to 4097 to reflect the initial amount of thermal photons in the mode (see Supplementary Information), which finally gave the experimental $\langle a^\dagger a\rangle$ plot in Figure~\ref{fig:dappc}(c). The set of coupled differential equations that govern the evolution of $\langle a^\dagger a\rangle$, $\langle a^\dagger \Tilde{S}^-\rangle$, $\langle \Tilde{S}^z\rangle$ and $\langle \Tilde{S}^+\Tilde{S}^-\rangle/N$ were solved using an explicit Runge-Kutta method of order 5(4) in Python (the default RK45 method of the `scipy.integrate.solve\_ivp' solver), and by choice of proper parameters, the simulated plot of $\langle a^\dagger a\rangle$ was fitted to the experimental plot, after which the resultant parameters were used for simulating the other three expectation values. The differential equations, initial conditions used for the simulation and fitted parameters are all listed in the Supplementary Information.


\section*{Acknowledgements}
We thank Ben Gaskell of Gaskell Quartz Ltd. (London) for making the strontium titanate ring used. We also thank Dr Ke-Jie Tan for growing DAP:PTP crystals many years ago, which helped when growing new DAP:PTP crystals for this work. We thank Professor Takeda Kazuyuki and Professor Yanai Nobuhiro for fruitful discussions on DAP:PTP and for lending samples for tests, as well as Dr Artem Bakulin for helpful discussions on the optical properties. This work was supported by the U.K. Engineering and Physical Sciences Research Council through Grant No. EP/V048430/1, EP/W027542/1 and EP/V001914/1. H.W. acknowledges financial support from the National Science Foundation of China (NSFC) (Grant No. 12204040) and the China Postdoctoral Science Foundation (Grant No. YJ20210035, No. 2021M700439).
\section*{Supplementary Information} 
Supplementary Information on the DFT spin-orbit coupling matrix elements, resonator coupling and cQED simulation parameters is available.
\newpage
\bibliography{Atemplate}

\begin{thebibliography}{32}%
\makeatletter
\providecommand \@ifxundefined [1]{%
 \@ifx{#1\undefined}
}%
\providecommand \@ifnum [1]{%
 \ifnum #1\expandafter \@firstoftwo
 \else \expandafter \@secondoftwo
 \fi
}%
\providecommand \@ifx [1]{%
 \ifx #1\expandafter \@firstoftwo
 \else \expandafter \@secondoftwo
 \fi
}%
\providecommand \natexlab [1]{#1}%
\providecommand \enquote  [1]{``#1''}%
\providecommand \bibnamefont  [1]{#1}%
\providecommand \bibfnamefont [1]{#1}%
\providecommand \citenamefont [1]{#1}%
\providecommand \href@noop [0]{\@secondoftwo}%
\providecommand \href [0]{\begingroup \@sanitize@url \@href}%
\providecommand \@href[1]{\@@startlink{#1}\@@href}%
\providecommand \@@href[1]{\endgroup#1\@@endlink}%
\providecommand \@sanitize@url [0]{\catcode `\\12\catcode `\$12\catcode
  `\&12\catcode `\#12\catcode `\^12\catcode `\_12\catcode `\%12\relax}%
\providecommand \@@startlink[1]{}%
\providecommand \@@endlink[0]{}%
\providecommand \url  [0]{\begingroup\@sanitize@url \@url }%
\providecommand \@url [1]{\endgroup\@href {#1}{\urlprefix }}%
\providecommand \urlprefix  [0]{URL }%
\providecommand \Eprint [0]{\href }%
\providecommand \doibase [0]{https://doi.org/}%
\providecommand \selectlanguage [0]{\@gobble}%
\providecommand \bibinfo  [0]{\@secondoftwo}%
\providecommand \bibfield  [0]{\@secondoftwo}%
\providecommand \translation [1]{[#1]}%
\providecommand \BibitemOpen [0]{}%
\providecommand \bibitemStop [0]{}%
\providecommand \bibitemNoStop [0]{.\EOS\space}%
\providecommand \EOS [0]{\spacefactor3000\relax}%
\providecommand \BibitemShut  [1]{\csname bibitem#1\endcsname}%
\let\auto@bib@innerbib\@empty
\bibitem [{\citenamefont {Breeze}\ \emph {et~al.}(2017)\citenamefont {Breeze},
  \citenamefont {Salvadori}, \citenamefont {Sathian}, \citenamefont {Alford},\
  and\ \citenamefont {Kay}}]{Breeze2017cqed}%
  \BibitemOpen
  \bibfield  {author} {\bibinfo {author} {\bibfnamefont {J.~D.}\ \bibnamefont
  {Breeze}}, \bibinfo {author} {\bibfnamefont {E.}~\bibnamefont {Salvadori}},
  \bibinfo {author} {\bibfnamefont {J.}~\bibnamefont {Sathian}}, \bibinfo
  {author} {\bibfnamefont {N.~M.}\ \bibnamefont {Alford}},\ and\ \bibinfo
  {author} {\bibfnamefont {C.~W.~M.}\ \bibnamefont {Kay}},\ }\bibfield  {title}
  {\bibinfo {title} {Room-temperature cavity quantum electrodynamics with
  strongly coupled dicke states},\ }\href
  {https://doi.org/10.1038/s41534-017-0041-3} {\bibfield  {journal} {\bibinfo
  {journal} {npj Quantum Inf.}\ }\textbf {\bibinfo {volume} {3}},\ \bibinfo
  {pages} {40} (\bibinfo {year} {2017})}\BibitemShut {NoStop}%
\bibitem [{\citenamefont {Zhang}\ \emph {et~al.}(2022)\citenamefont {Zhang},
  \citenamefont {Wu}, \citenamefont {Su}, \citenamefont {Lou}, \citenamefont
  {Shan},\ and\ \citenamefont {M\o{}lmer}}]{Zhang2022}%
  \BibitemOpen
  \bibfield  {author} {\bibinfo {author} {\bibfnamefont {Y.}~\bibnamefont
  {Zhang}}, \bibinfo {author} {\bibfnamefont {Q.}~\bibnamefont {Wu}}, \bibinfo
  {author} {\bibfnamefont {S.-L.}\ \bibnamefont {Su}}, \bibinfo {author}
  {\bibfnamefont {Q.}~\bibnamefont {Lou}}, \bibinfo {author} {\bibfnamefont
  {C.}~\bibnamefont {Shan}},\ and\ \bibinfo {author} {\bibfnamefont
  {K.}~\bibnamefont {M\o{}lmer}},\ }\bibfield  {title} {\bibinfo {title}
  {Cavity quantum electrodynamics effects with nitrogen vacancy center spins
  coupled to room temperature microwave resonators},\ }\href
  {https://doi.org/10.1103/PhysRevLett.128.253601} {\bibfield  {journal}
  {\bibinfo  {journal} {Phys. Rev. Lett.}\ }\textbf {\bibinfo {volume} {128}},\
  \bibinfo {pages} {253601} (\bibinfo {year} {2022})}\BibitemShut {NoStop}%
\bibitem [{\citenamefont {Oxborrow}\ \emph {et~al.}(2012)\citenamefont
  {Oxborrow}, \citenamefont {Breeze},\ and\ \citenamefont
  {Alford}}]{Oxborrow2012}%
  \BibitemOpen
  \bibfield  {author} {\bibinfo {author} {\bibfnamefont {M.}~\bibnamefont
  {Oxborrow}}, \bibinfo {author} {\bibfnamefont {J.~D.}\ \bibnamefont
  {Breeze}},\ and\ \bibinfo {author} {\bibfnamefont {N.~M.}\ \bibnamefont
  {Alford}},\ }\bibfield  {title} {\bibinfo {title} {{Room-temperature
  solid-state maser}},\ }\href {https://doi.org/10.1038/nature11339} {\bibfield
   {journal} {\bibinfo  {journal} {Nature}\ }\textbf {\bibinfo {volume}
  {488}},\ \bibinfo {pages} {353} (\bibinfo {year} {2012})}\BibitemShut
  {NoStop}%
\bibitem [{\citenamefont {Breeze}\ \emph {et~al.}(2018)\citenamefont {Breeze},
  \citenamefont {Salvadori}, \citenamefont {Sathian}, \citenamefont {Alford},\
  and\ \citenamefont {Kay}}]{Breeze2018}%
  \BibitemOpen
  \bibfield  {author} {\bibinfo {author} {\bibfnamefont {J.}~\bibnamefont
  {Breeze}}, \bibinfo {author} {\bibfnamefont {E.}~\bibnamefont {Salvadori}},
  \bibinfo {author} {\bibfnamefont {J.}~\bibnamefont {Sathian}}, \bibinfo
  {author} {\bibfnamefont {N.~M.}\ \bibnamefont {Alford}},\ and\ \bibinfo
  {author} {\bibfnamefont {C.~W.~M.}\ \bibnamefont {Kay}},\ }\bibfield  {title}
  {\bibinfo {title} {Continuous-wave room-temperature diamond maser},\
  }\href@noop {} {\bibfield  {journal} {\bibinfo  {journal} {Nature}\ }\textbf
  {\bibinfo {volume} {555}},\ \bibinfo {pages} {493} (\bibinfo {year}
  {2018})}\BibitemShut {NoStop}%
\bibitem [{\citenamefont {Bogatko}\ \emph {et~al.}(2016)\citenamefont
  {Bogatko}, \citenamefont {Haynes}, \citenamefont {Sathian}, \citenamefont
  {Wade}, \citenamefont {Kim}, \citenamefont {Tan}, \citenamefont {Breeze},
  \citenamefont {Salvadori}, \citenamefont {Horsfield},\ and\ \citenamefont
  {Oxborrow}}]{Bogatko2016}%
  \BibitemOpen
  \bibfield  {author} {\bibinfo {author} {\bibfnamefont {S.}~\bibnamefont
  {Bogatko}}, \bibinfo {author} {\bibfnamefont {P.~D.}\ \bibnamefont {Haynes}},
  \bibinfo {author} {\bibfnamefont {J.}~\bibnamefont {Sathian}}, \bibinfo
  {author} {\bibfnamefont {J.}~\bibnamefont {Wade}}, \bibinfo {author}
  {\bibfnamefont {J.~S.}\ \bibnamefont {Kim}}, \bibinfo {author} {\bibfnamefont
  {K.~J.}\ \bibnamefont {Tan}}, \bibinfo {author} {\bibfnamefont
  {J.}~\bibnamefont {Breeze}}, \bibinfo {author} {\bibfnamefont
  {E.}~\bibnamefont {Salvadori}}, \bibinfo {author} {\bibfnamefont
  {A.}~\bibnamefont {Horsfield}},\ and\ \bibinfo {author} {\bibfnamefont
  {M.}~\bibnamefont {Oxborrow}},\ }\bibfield  {title} {\bibinfo {title}
  {{Molecular Design of a Room-Temperature Maser}},\ }\href
  {https://doi.org/10.1021/acs.jpcc.6b00150} {\bibfield  {journal} {\bibinfo
  {journal} {J. Phys. Chem. C}\ }\textbf {\bibinfo {volume} {120}},\ \bibinfo
  {pages} {8251} (\bibinfo {year} {2016})}\BibitemShut {NoStop}%
\bibitem [{\citenamefont {Patterson}\ \emph {et~al.}(1984)\citenamefont
  {Patterson}, \citenamefont {Lee}, \citenamefont {Wilson},\ and\ \citenamefont
  {Fayer}}]{Patterson1984}%
  \BibitemOpen
  \bibfield  {author} {\bibinfo {author} {\bibfnamefont {F.}~\bibnamefont
  {Patterson}}, \bibinfo {author} {\bibfnamefont {H.}~\bibnamefont {Lee}},
  \bibinfo {author} {\bibfnamefont {W.~L.}\ \bibnamefont {Wilson}},\ and\
  \bibinfo {author} {\bibfnamefont {M.}~\bibnamefont {Fayer}},\ }\bibfield
  {title} {\bibinfo {title} {Intersystem crossing from singlet states of
  molecular dimers and monomers in mixed molecular crystals: picosecond
  stimulated photon echo experiments},\ }\href
  {https://doi.org/https://doi.org/10.1016/0301-0104(84)80005-1} {\bibfield
  {journal} {\bibinfo  {journal} {Chem. Phys.}\ }\textbf {\bibinfo {volume}
  {84}},\ \bibinfo {pages} {51} (\bibinfo {year} {1984})}\BibitemShut {NoStop}%
\bibitem [{\citenamefont {Wang}\ \emph {et~al.}(2022)\citenamefont {Wang},
  \citenamefont {Miao}, \citenamefont {Cai}, \citenamefont {Ding},
  \citenamefont {Li}, \citenamefont {Li}, \citenamefont {Zhu}, \citenamefont
  {Tao}, \citenamefont {Jia}, \citenamefont {Liang}, \citenamefont {Lu},
  \citenamefont {Fang}, \citenamefont {Yi},\ and\ \citenamefont
  {Lin}}]{Wang2022}%
  \BibitemOpen
  \bibfield  {author} {\bibinfo {author} {\bibfnamefont {W.}~\bibnamefont
  {Wang}}, \bibinfo {author} {\bibfnamefont {X.}~\bibnamefont {Miao}}, \bibinfo
  {author} {\bibfnamefont {G.}~\bibnamefont {Cai}}, \bibinfo {author}
  {\bibfnamefont {L.}~\bibnamefont {Ding}}, \bibinfo {author} {\bibfnamefont
  {Y.}~\bibnamefont {Li}}, \bibinfo {author} {\bibfnamefont {T.}~\bibnamefont
  {Li}}, \bibinfo {author} {\bibfnamefont {Y.}~\bibnamefont {Zhu}}, \bibinfo
  {author} {\bibfnamefont {L.}~\bibnamefont {Tao}}, \bibinfo {author}
  {\bibfnamefont {Y.}~\bibnamefont {Jia}}, \bibinfo {author} {\bibfnamefont
  {Y.}~\bibnamefont {Liang}}, \bibinfo {author} {\bibfnamefont
  {X.}~\bibnamefont {Lu}}, \bibinfo {author} {\bibfnamefont {Y.}~\bibnamefont
  {Fang}}, \bibinfo {author} {\bibfnamefont {Y.}~\bibnamefont {Yi}},\ and\
  \bibinfo {author} {\bibfnamefont {Y.}~\bibnamefont {Lin}},\ }\bibfield
  {title} {\bibinfo {title} {Enhancing transition dipole moments of
  heterocyclic semiconductors via rational nitrogen-substitution for sensitive
  near infrared detection},\ }\href@noop {} {\bibfield  {journal} {\bibinfo
  {journal} {Adv. Mater.}\ }\textbf {\bibinfo {volume} {34}},\ \bibinfo {pages}
  {2201600} (\bibinfo {year} {2022})}\BibitemShut {NoStop}%
\bibitem [{\citenamefont {de~Souza}\ \emph {et~al.}(2016)\citenamefont
  {de~Souza}, \citenamefont {Vivas}, \citenamefont {Mendon\c{c}a},
  \citenamefont {Plunkett}, \citenamefont {Filatov}, \citenamefont {Senge},\
  and\ \citenamefont {De~Boni}}]{deSouza2016}%
  \BibitemOpen
  \bibfield  {author} {\bibinfo {author} {\bibfnamefont {T.~G.~B.}\
  \bibnamefont {de~Souza}}, \bibinfo {author} {\bibfnamefont {M.~G.}\
  \bibnamefont {Vivas}}, \bibinfo {author} {\bibfnamefont {C.~R.}\ \bibnamefont
  {Mendon\c{c}a}}, \bibinfo {author} {\bibfnamefont {S.}~\bibnamefont
  {Plunkett}}, \bibinfo {author} {\bibfnamefont {M.~A.}\ \bibnamefont
  {Filatov}}, \bibinfo {author} {\bibfnamefont {M.~O.}\ \bibnamefont {Senge}},\
  and\ \bibinfo {author} {\bibfnamefont {L.}~\bibnamefont {De~Boni}},\
  }\bibfield  {title} {\bibinfo {title} {Studying the intersystem crossing rate
  and triplet quantum yield of meso-substituted porphyrins by means of pulse
  train fluorescence technique},\ }\href
  {https://doi.org/10.1142/S1088424616500048} {\bibfield  {journal} {\bibinfo
  {journal} {J. Porphyrins Phthalocyanines}\ }\textbf {\bibinfo {volume}
  {20}},\ \bibinfo {pages} {282} (\bibinfo {year} {2016})}\BibitemShut
  {NoStop}%
\bibitem [{\citenamefont {Takeda}\ \emph {et~al.}(2002)\citenamefont {Takeda},
  \citenamefont {Takegoshi},\ and\ \citenamefont {Terao}}]{Takeda2002}%
  \BibitemOpen
  \bibfield  {author} {\bibinfo {author} {\bibfnamefont {K.}~\bibnamefont
  {Takeda}}, \bibinfo {author} {\bibfnamefont {K.}~\bibnamefont {Takegoshi}},\
  and\ \bibinfo {author} {\bibfnamefont {T.}~\bibnamefont {Terao}},\ }\bibfield
   {title} {\bibinfo {title} {Zero-field electron spin resonance and
  theoretical studies of light penetration into single crystal and
  polycrystalline material doped with molecules photoexcitable to the triplet
  state via intersystem crossing},\ }\href {https://doi.org/10.1063/1.1499124}
  {\bibfield  {journal} {\bibinfo  {journal} {J. Chem. Phys.}\ }\textbf
  {\bibinfo {volume} {117}},\ \bibinfo {pages} {4940} (\bibinfo {year}
  {2002})}\BibitemShut {NoStop}%
\bibitem [{\citenamefont {Kryschi}\ \emph {et~al.}(1992)\citenamefont
  {Kryschi}, \citenamefont {Krüger}, \citenamefont {Wagner},\ and\
  \citenamefont {Gorgas}}]{Kryschi1992}%
  \BibitemOpen
  \bibfield  {author} {\bibinfo {author} {\bibfnamefont {C.}~\bibnamefont
  {Kryschi}}, \bibinfo {author} {\bibfnamefont {A.}~\bibnamefont {Krüger}},
  \bibinfo {author} {\bibfnamefont {B.}~\bibnamefont {Wagner}},\ and\ \bibinfo
  {author} {\bibfnamefont {W.}~\bibnamefont {Gorgas}},\ }\bibfield  {title}
  {\bibinfo {title} {Intermolecular and intramolecular radiationless processes
  in p-terphenyl and benzoic-acid single crystals doped with tetra-cene or
  pentacene},\ }\href {https://doi.org/10.1080/10587259208047007} {\bibfield
  {journal} {\bibinfo  {journal} {Mol. Cryst. Liq. Cryst. Sci. Technol., Sect.
  A}\ }\textbf {\bibinfo {volume} {218}},\ \bibinfo {pages} {7} (\bibinfo
  {year} {1992})}\BibitemShut {NoStop}%
\bibitem [{\citenamefont {Williams}\ \emph {et~al.}(1983)\citenamefont
  {Williams}, \citenamefont {Jones},\ and\ \citenamefont
  {Davies}}]{Williams1983}%
  \BibitemOpen
  \bibfield  {author} {\bibinfo {author} {\bibfnamefont {J.~O.}\ \bibnamefont
  {Williams}}, \bibinfo {author} {\bibfnamefont {A.~C.}\ \bibnamefont
  {Jones}},\ and\ \bibinfo {author} {\bibfnamefont {M.~J.}\ \bibnamefont
  {Davies}},\ }\bibfield  {title} {\bibinfo {title} {Radiationless transitions
  in p-terphenyl crystals doped with anthracene{,} tetracene and pentacene},\
  }\href {https://doi.org/10.1039/F29837900263} {\bibfield  {journal} {\bibinfo
   {journal} {J. Chem. Soc.{,} Faraday Trans. 2}\ }\textbf {\bibinfo {volume}
  {79}},\ \bibinfo {pages} {263} (\bibinfo {year} {1983})}\BibitemShut
  {NoStop}%
\bibitem [{\citenamefont {Brouwer}\ \emph {et~al.}(1999)\citenamefont
  {Brouwer}, \citenamefont {Köhler}, \citenamefont {van Oijen}, \citenamefont
  {Groenen},\ and\ \citenamefont {Schmidt}}]{Brouwer1999}%
  \BibitemOpen
  \bibfield  {author} {\bibinfo {author} {\bibfnamefont {A.~C.~J.}\
  \bibnamefont {Brouwer}}, \bibinfo {author} {\bibfnamefont {J.}~\bibnamefont
  {Köhler}}, \bibinfo {author} {\bibfnamefont {A.~M.}\ \bibnamefont {van
  Oijen}}, \bibinfo {author} {\bibfnamefont {E.~J.~J.}\ \bibnamefont
  {Groenen}},\ and\ \bibinfo {author} {\bibfnamefont {J.}~\bibnamefont
  {Schmidt}},\ }\bibfield  {title} {\bibinfo {title} {Single-molecule
  fluorescence autocorrelation experiments on pentacene: The dependence of
  intersystem crossing on isotopic composition},\ }\href
  {https://doi.org/10.1063/1.478837} {\bibfield  {journal} {\bibinfo  {journal}
  {J. Chem. Phys.}\ }\textbf {\bibinfo {volume} {110}},\ \bibinfo {pages}
  {9151} (\bibinfo {year} {1999})}\BibitemShut {NoStop}%
\bibitem [{\citenamefont {Kouno}\ \emph {et~al.}(2019)\citenamefont {Kouno},
  \citenamefont {Kawashima}, \citenamefont {Tateishi}, \citenamefont {Uesaka},
  \citenamefont {Kimizuka},\ and\ \citenamefont {Yanai}}]{kouno2019}%
  \BibitemOpen
  \bibfield  {author} {\bibinfo {author} {\bibfnamefont {H.}~\bibnamefont
  {Kouno}}, \bibinfo {author} {\bibfnamefont {Y.}~\bibnamefont {Kawashima}},
  \bibinfo {author} {\bibfnamefont {K.}~\bibnamefont {Tateishi}}, \bibinfo
  {author} {\bibfnamefont {T.}~\bibnamefont {Uesaka}}, \bibinfo {author}
  {\bibfnamefont {N.}~\bibnamefont {Kimizuka}},\ and\ \bibinfo {author}
  {\bibfnamefont {N.}~\bibnamefont {Yanai}},\ }\bibfield  {title} {\bibinfo
  {title} {Nonpentacene polarizing agents with improved air stability for
  triplet dynamic nuclear polarization at room temperature},\ }\href@noop {}
  {\bibfield  {journal} {\bibinfo  {journal} {J. Phys. Chem. Lett.}\ }\textbf
  {\bibinfo {volume} {10}},\ \bibinfo {pages} {2208} (\bibinfo {year}
  {2019})}\BibitemShut {NoStop}%
\bibitem [{\citenamefont {Pedash}\ \emph {et~al.}(2002)\citenamefont {Pedash},
  \citenamefont {Prezhdo}, \citenamefont {Kotelevskiy},\ and\ \citenamefont
  {Prezhdo}}]{Pedash2002}%
  \BibitemOpen
  \bibfield  {author} {\bibinfo {author} {\bibfnamefont {Y.}~\bibnamefont
  {Pedash}}, \bibinfo {author} {\bibfnamefont {O.}~\bibnamefont {Prezhdo}},
  \bibinfo {author} {\bibfnamefont {S.}~\bibnamefont {Kotelevskiy}},\ and\
  \bibinfo {author} {\bibfnamefont {V.}~\bibnamefont {Prezhdo}},\ }\bibfield
  {title} {\bibinfo {title} {Spin–orbit coupling and luminescence
  characteristics of conjugated organic molecules. i. polyacenes},\ }\href
  {https://doi.org/https://doi.org/10.1016/S0166-1280(02)00035-0} {\bibfield
  {journal} {\bibinfo  {journal} {J. Mol. Struct.: THEOCHEM}\ }\textbf
  {\bibinfo {volume} {585}},\ \bibinfo {pages} {49} (\bibinfo {year}
  {2002})}\BibitemShut {NoStop}%
\bibitem [{\citenamefont {Schott}\ \emph {et~al.}(2017)\citenamefont {Schott},
  \citenamefont {McNellis}, \citenamefont {Nielsen}, \citenamefont {Chen},
  \citenamefont {Watanabe}, \citenamefont {Tanaka}, \citenamefont {McCulloch},
  \citenamefont {Takimiya}, \citenamefont {Sinova},\ and\ \citenamefont
  {Sirringhaus}}]{Schott2017}%
  \BibitemOpen
  \bibfield  {author} {\bibinfo {author} {\bibfnamefont {S.}~\bibnamefont
  {Schott}}, \bibinfo {author} {\bibfnamefont {E.~R.}\ \bibnamefont
  {McNellis}}, \bibinfo {author} {\bibfnamefont {C.~B.}\ \bibnamefont
  {Nielsen}}, \bibinfo {author} {\bibfnamefont {H.-Y.}\ \bibnamefont {Chen}},
  \bibinfo {author} {\bibfnamefont {S.}~\bibnamefont {Watanabe}}, \bibinfo
  {author} {\bibfnamefont {H.}~\bibnamefont {Tanaka}}, \bibinfo {author}
  {\bibfnamefont {I.}~\bibnamefont {McCulloch}}, \bibinfo {author}
  {\bibfnamefont {K.}~\bibnamefont {Takimiya}}, \bibinfo {author}
  {\bibfnamefont {J.}~\bibnamefont {Sinova}},\ and\ \bibinfo {author}
  {\bibfnamefont {H.}~\bibnamefont {Sirringhaus}},\ }\bibfield  {title}
  {\bibinfo {title} {Tuning the effective spin-orbit coupling in molecular
  semiconductors},\ }\href {https://doi.org/10.1038/ncomms15200} {\bibfield
  {journal} {\bibinfo  {journal} {Nat. Commun.}\ }\textbf {\bibinfo {volume}
  {8}},\ \bibinfo {pages} {15200} (\bibinfo {year} {2017})}\BibitemShut
  {NoStop}%
\bibitem [{\citenamefont {Marian}(2021)}]{Marian2021}%
  \BibitemOpen
  \bibfield  {author} {\bibinfo {author} {\bibfnamefont {C.~M.}\ \bibnamefont
  {Marian}},\ }\bibfield  {title} {\bibinfo {title} {Understanding and
  controlling intersystem crossing in molecules},\ }\href
  {https://doi.org/10.1146/annurev-physchem-061020-053433} {\bibfield
  {journal} {\bibinfo  {journal} {Annu. Rev. Phys. Chem.}\ }\textbf {\bibinfo
  {volume} {72}},\ \bibinfo {pages} {617} (\bibinfo {year} {2021})}\BibitemShut
  {NoStop}%
\bibitem [{\citenamefont {Charlton}\ \emph {et~al.}(2018)\citenamefont
  {Charlton}, \citenamefont {Fogarty}, \citenamefont {Bogatko}, \citenamefont
  {Zuehlsdorff}, \citenamefont {Hine}, \citenamefont {Heeney}, \citenamefont
  {Horsfield},\ and\ \citenamefont {Haynes}}]{Charlton2018}%
  \BibitemOpen
  \bibfield  {author} {\bibinfo {author} {\bibfnamefont {R.~J.}\ \bibnamefont
  {Charlton}}, \bibinfo {author} {\bibfnamefont {R.~M.}\ \bibnamefont
  {Fogarty}}, \bibinfo {author} {\bibfnamefont {S.}~\bibnamefont {Bogatko}},
  \bibinfo {author} {\bibfnamefont {T.~J.}\ \bibnamefont {Zuehlsdorff}},
  \bibinfo {author} {\bibfnamefont {N.~D.~M.}\ \bibnamefont {Hine}}, \bibinfo
  {author} {\bibfnamefont {M.}~\bibnamefont {Heeney}}, \bibinfo {author}
  {\bibfnamefont {A.~P.}\ \bibnamefont {Horsfield}},\ and\ \bibinfo {author}
  {\bibfnamefont {P.~D.}\ \bibnamefont {Haynes}},\ }\bibfield  {title}
  {\bibinfo {title} {Implicit and explicit host effects on excitons in
  pentacene derivatives},\ }\href {https://doi.org/10.1063/1.5017285}
  {\bibfield  {journal} {\bibinfo  {journal} {J. Chem. Phys.}\ }\textbf
  {\bibinfo {volume} {148}},\ \bibinfo {pages} {104108} (\bibinfo {year}
  {2018})}\BibitemShut {NoStop}%
\bibitem [{\citenamefont {Bertoni}\ \emph {et~al.}(2022)\citenamefont
  {Bertoni}, \citenamefont {Fogarty}, \citenamefont {Sánchez},\ and\
  \citenamefont {Horsfield}}]{Bertoni2022}%
  \BibitemOpen
  \bibfield  {author} {\bibinfo {author} {\bibfnamefont {A.~I.}\ \bibnamefont
  {Bertoni}}, \bibinfo {author} {\bibfnamefont {R.~M.}\ \bibnamefont
  {Fogarty}}, \bibinfo {author} {\bibfnamefont {C.~G.}\ \bibnamefont
  {Sánchez}},\ and\ \bibinfo {author} {\bibfnamefont {A.~P.}\ \bibnamefont
  {Horsfield}},\ }\bibfield  {title} {\bibinfo {title} {Qm/mm optimization with
  quantum coupling: Host–guest interactions in a pentacene-doped p-terphenyl
  crystal},\ }\href {https://doi.org/10.1063/5.0079788} {\bibfield  {journal}
  {\bibinfo  {journal} {J. Chem. Phys.}\ }\textbf {\bibinfo {volume} {156}},\
  \bibinfo {pages} {044110} (\bibinfo {year} {2022})}\BibitemShut {NoStop}%
\bibitem [{\citenamefont {El‐Sayed}(1963)}]{ElSayed63}%
  \BibitemOpen
  \bibfield  {author} {\bibinfo {author} {\bibfnamefont {M.~A.}\ \bibnamefont
  {El‐Sayed}},\ }\bibfield  {title} {\bibinfo {title} {Spin—orbit coupling
  and the radiationless processes in nitrogen heterocyclics},\ }\href
  {https://doi.org/10.1063/1.1733610} {\bibfield  {journal} {\bibinfo
  {journal} {J. Chem. Phys.}\ }\textbf {\bibinfo {volume} {38}},\ \bibinfo
  {pages} {2834} (\bibinfo {year} {1963})}\BibitemShut {NoStop}%
\bibitem [{\citenamefont {Matsuoka}\ \emph {et~al.}(2017)\citenamefont
  {Matsuoka}, \citenamefont {Retegan}, \citenamefont {Schmitt}, \citenamefont
  {Höger}, \citenamefont {Neese},\ and\ \citenamefont
  {Schiemann}}]{Matsuoka_biexp}%
  \BibitemOpen
  \bibfield  {author} {\bibinfo {author} {\bibfnamefont {H.}~\bibnamefont
  {Matsuoka}}, \bibinfo {author} {\bibfnamefont {M.}~\bibnamefont {Retegan}},
  \bibinfo {author} {\bibfnamefont {L.}~\bibnamefont {Schmitt}}, \bibinfo
  {author} {\bibfnamefont {S.}~\bibnamefont {Höger}}, \bibinfo {author}
  {\bibfnamefont {F.}~\bibnamefont {Neese}},\ and\ \bibinfo {author}
  {\bibfnamefont {O.}~\bibnamefont {Schiemann}},\ }\bibfield  {title} {\bibinfo
  {title} {Time-resolved electron paramagnetic resonance and theoretical
  investigations of metal-free room-temperature triplet emitters},\ }\href
  {https://doi.org/10.1021/jacs.7b04561} {\bibfield  {journal} {\bibinfo
  {journal} {J. Am. Chem. Soc.}\ }\textbf {\bibinfo {volume} {139}},\ \bibinfo
  {pages} {12968} (\bibinfo {year} {2017})}\BibitemShut {NoStop}%
\bibitem [{\citenamefont {Hintze}\ \emph {et~al.}(2017)\citenamefont {Hintze},
  \citenamefont {Steiner},\ and\ \citenamefont {Drescher}}]{Hintze2017}%
  \BibitemOpen
  \bibfield  {author} {\bibinfo {author} {\bibfnamefont {C.}~\bibnamefont
  {Hintze}}, \bibinfo {author} {\bibfnamefont {U.~E.}\ \bibnamefont
  {Steiner}},\ and\ \bibinfo {author} {\bibfnamefont {M.}~\bibnamefont
  {Drescher}},\ }\bibfield  {title} {\bibinfo {title} {Photoexcited triplet
  state kinetics studied by electron paramagnetic resonance spectroscopy},\
  }\href {https://doi.org/10.1002/cphc.201600868} {\bibfield  {journal}
  {\bibinfo  {journal} {ChemPhysChem}\ }\textbf {\bibinfo {volume} {18}},\
  \bibinfo {pages} {6} (\bibinfo {year} {2017})}\BibitemShut {NoStop}%
\bibitem [{\citenamefont {Breeze}\ \emph {et~al.}(2015)\citenamefont {Breeze},
  \citenamefont {Tan}, \citenamefont {Richards}, \citenamefont {Sathian},
  \citenamefont {Oxborrow},\ and\ \citenamefont {Alford}}]{breeze2015enhanced}%
  \BibitemOpen
  \bibfield  {author} {\bibinfo {author} {\bibfnamefont {J.}~\bibnamefont
  {Breeze}}, \bibinfo {author} {\bibfnamefont {K.-J.}\ \bibnamefont {Tan}},
  \bibinfo {author} {\bibfnamefont {B.}~\bibnamefont {Richards}}, \bibinfo
  {author} {\bibfnamefont {J.}~\bibnamefont {Sathian}}, \bibinfo {author}
  {\bibfnamefont {M.}~\bibnamefont {Oxborrow}},\ and\ \bibinfo {author}
  {\bibfnamefont {N.~M.}\ \bibnamefont {Alford}},\ }\bibfield  {title}
  {\bibinfo {title} {Enhanced magnetic purcell effect in room-temperature
  masers},\ }\href@noop {} {\bibfield  {journal} {\bibinfo  {journal} {Nat.
  Commun.}\ }\textbf {\bibinfo {volume} {6}},\ \bibinfo {pages} {1} (\bibinfo
  {year} {2015})}\BibitemShut {NoStop}%
\bibitem [{\citenamefont {Wu}\ \emph {et~al.}(2019)\citenamefont {Wu},
  \citenamefont {Ng}, \citenamefont {Mirkhanov}, \citenamefont {Amirzhan},
  \citenamefont {Nitnara},\ and\ \citenamefont {Oxborrow}}]{HaoWu2019}%
  \BibitemOpen
  \bibfield  {author} {\bibinfo {author} {\bibfnamefont {H.}~\bibnamefont
  {Wu}}, \bibinfo {author} {\bibfnamefont {W.}~\bibnamefont {Ng}}, \bibinfo
  {author} {\bibfnamefont {S.}~\bibnamefont {Mirkhanov}}, \bibinfo {author}
  {\bibfnamefont {A.}~\bibnamefont {Amirzhan}}, \bibinfo {author}
  {\bibfnamefont {S.}~\bibnamefont {Nitnara}},\ and\ \bibinfo {author}
  {\bibfnamefont {M.}~\bibnamefont {Oxborrow}},\ }\bibfield  {title} {\bibinfo
  {title} {Unraveling the room-temperature spin dynamics of photoexcited
  pentacene in its lowest triplet state at zero field},\ }\href
  {https://doi.org/10.1021/acs.jpcc.9b08439} {\bibfield  {journal} {\bibinfo
  {journal} {J. Phys. Chem. C}\ }\textbf {\bibinfo {volume} {123}},\ \bibinfo
  {pages} {24275} (\bibinfo {year} {2019})}\BibitemShut {NoStop}%
\bibitem [{\citenamefont {Kajfez}(1995)}]{kajfez1995}%
  \BibitemOpen
  \bibfield  {author} {\bibinfo {author} {\bibfnamefont {D.}~\bibnamefont
  {Kajfez}},\ }\bibfield  {title} {\bibinfo {title} {Q-factor measurement with
  a scalar network analyser},\ }\href
  {https://digital-library.theiet.org/content/journals/10.1049/ip-map_19952142}
  {\bibfield  {journal} {\bibinfo  {journal} {IEE Proceedings - Microwaves,
  Antennas and Propagation}\ }\textbf {\bibinfo {volume} {142}},\ \bibinfo
  {pages} {369} (\bibinfo {year} {1995})}\BibitemShut {NoStop}%
\bibitem [{\citenamefont {Kajfez}()}]{kajfezwebsite}%
  \BibitemOpen
  \bibfield  {author} {\bibinfo {author} {\bibfnamefont {D.}~\bibnamefont
  {Kajfez}},\ }\href@noop {} {\bibinfo {title} {Q factor measurements, analog
  and digital}},\ \bibinfo {howpublished} {Available:
  https://people.engineering.olemiss.edu/darko-kajfez/assets/rfqmeas2b.pdf},\
  \bibinfo {note} {(Accessed: 2022-07-30)}\BibitemShut {NoStop}%
\bibitem [{\citenamefont {Eisenach}\ \emph {et~al.}(2021)\citenamefont
  {Eisenach}, \citenamefont {Barry}, \citenamefont {O'Keeffe}, \citenamefont
  {Schloss}, \citenamefont {Steinecker}, \citenamefont {Englund},\ and\
  \citenamefont {Braje}}]{Eisenach2021}%
  \BibitemOpen
  \bibfield  {author} {\bibinfo {author} {\bibfnamefont {E.~R.}\ \bibnamefont
  {Eisenach}}, \bibinfo {author} {\bibfnamefont {J.~F.}\ \bibnamefont {Barry}},
  \bibinfo {author} {\bibfnamefont {M.~F.}\ \bibnamefont {O'Keeffe}}, \bibinfo
  {author} {\bibfnamefont {J.~M.}\ \bibnamefont {Schloss}}, \bibinfo {author}
  {\bibfnamefont {M.~H.}\ \bibnamefont {Steinecker}}, \bibinfo {author}
  {\bibfnamefont {D.~R.}\ \bibnamefont {Englund}},\ and\ \bibinfo {author}
  {\bibfnamefont {D.~A.}\ \bibnamefont {Braje}},\ }\bibfield  {title} {\bibinfo
  {title} {Cavity-enhanced microwave readout of a solid-state spin sensor},\
  }\href {https://doi.org/10.1038/s41467-021-21256-7} {\bibfield  {journal}
  {\bibinfo  {journal} {Nat. Commun.}\ }\textbf {\bibinfo {volume} {12}},\
  \bibinfo {pages} {1357} (\bibinfo {year} {2021})}\BibitemShut {NoStop}%
\bibitem [{\citenamefont {Ams\"uss}\ \emph {et~al.}(2011)\citenamefont
  {Ams\"uss}, \citenamefont {Koller}, \citenamefont {N\"obauer}, \citenamefont
  {Putz}, \citenamefont {Rotter}, \citenamefont {Sandner}, \citenamefont
  {Schneider}, \citenamefont {Schramb\"ock}, \citenamefont {Steinhauser},
  \citenamefont {Ritsch}, \citenamefont {Schmiedmayer},\ and\ \citenamefont
  {Majer}}]{Amsuss2011}%
  \BibitemOpen
  \bibfield  {author} {\bibinfo {author} {\bibfnamefont {R.}~\bibnamefont
  {Ams\"uss}}, \bibinfo {author} {\bibfnamefont {C.}~\bibnamefont {Koller}},
  \bibinfo {author} {\bibfnamefont {T.}~\bibnamefont {N\"obauer}}, \bibinfo
  {author} {\bibfnamefont {S.}~\bibnamefont {Putz}}, \bibinfo {author}
  {\bibfnamefont {S.}~\bibnamefont {Rotter}}, \bibinfo {author} {\bibfnamefont
  {K.}~\bibnamefont {Sandner}}, \bibinfo {author} {\bibfnamefont
  {S.}~\bibnamefont {Schneider}}, \bibinfo {author} {\bibfnamefont
  {M.}~\bibnamefont {Schramb\"ock}}, \bibinfo {author} {\bibfnamefont
  {G.}~\bibnamefont {Steinhauser}}, \bibinfo {author} {\bibfnamefont
  {H.}~\bibnamefont {Ritsch}}, \bibinfo {author} {\bibfnamefont
  {J.}~\bibnamefont {Schmiedmayer}},\ and\ \bibinfo {author} {\bibfnamefont
  {J.}~\bibnamefont {Majer}},\ }\bibfield  {title} {\bibinfo {title} {Cavity
  {QED} with magnetically coupled collective spin states},\ }\href
  {https://doi.org/10.1103/PhysRevLett.107.060502} {\bibfield  {journal}
  {\bibinfo  {journal} {Phys. Rev. Lett.}\ }\textbf {\bibinfo {volume} {107}},\
  \bibinfo {pages} {060502} (\bibinfo {year} {2011})}\BibitemShut {NoStop}%
\bibitem [{\citenamefont {Kubo}\ \emph {et~al.}(2011)\citenamefont {Kubo},
  \citenamefont {Grezes}, \citenamefont {Dewes}, \citenamefont {Umeda},
  \citenamefont {Isoya}, \citenamefont {Sumiya}, \citenamefont {Morishita},
  \citenamefont {Abe}, \citenamefont {Onoda}, \citenamefont {Ohshima},
  \citenamefont {Jacques}, \citenamefont {Dr\'eau}, \citenamefont {Roch},
  \citenamefont {Diniz}, \citenamefont {Auffeves}, \citenamefont {Vion},
  \citenamefont {Esteve},\ and\ \citenamefont {Bertet}}]{Kubo2011}%
  \BibitemOpen
  \bibfield  {author} {\bibinfo {author} {\bibfnamefont {Y.}~\bibnamefont
  {Kubo}}, \bibinfo {author} {\bibfnamefont {C.}~\bibnamefont {Grezes}},
  \bibinfo {author} {\bibfnamefont {A.}~\bibnamefont {Dewes}}, \bibinfo
  {author} {\bibfnamefont {T.}~\bibnamefont {Umeda}}, \bibinfo {author}
  {\bibfnamefont {J.}~\bibnamefont {Isoya}}, \bibinfo {author} {\bibfnamefont
  {H.}~\bibnamefont {Sumiya}}, \bibinfo {author} {\bibfnamefont
  {N.}~\bibnamefont {Morishita}}, \bibinfo {author} {\bibfnamefont
  {H.}~\bibnamefont {Abe}}, \bibinfo {author} {\bibfnamefont {S.}~\bibnamefont
  {Onoda}}, \bibinfo {author} {\bibfnamefont {T.}~\bibnamefont {Ohshima}},
  \bibinfo {author} {\bibfnamefont {V.}~\bibnamefont {Jacques}}, \bibinfo
  {author} {\bibfnamefont {A.}~\bibnamefont {Dr\'eau}}, \bibinfo {author}
  {\bibfnamefont {J.-F.}\ \bibnamefont {Roch}}, \bibinfo {author}
  {\bibfnamefont {I.}~\bibnamefont {Diniz}}, \bibinfo {author} {\bibfnamefont
  {A.}~\bibnamefont {Auffeves}}, \bibinfo {author} {\bibfnamefont
  {D.}~\bibnamefont {Vion}}, \bibinfo {author} {\bibfnamefont {D.}~\bibnamefont
  {Esteve}},\ and\ \bibinfo {author} {\bibfnamefont {P.}~\bibnamefont
  {Bertet}},\ }\bibfield  {title} {\bibinfo {title} {Hybrid quantum circuit
  with a superconducting qubit coupled to a spin ensemble},\ }\href
  {https://doi.org/10.1103/PhysRevLett.107.220501} {\bibfield  {journal}
  {\bibinfo  {journal} {Phys. Rev. Lett.}\ }\textbf {\bibinfo {volume} {107}},\
  \bibinfo {pages} {220501} (\bibinfo {year} {2011})}\BibitemShut {NoStop}%
\bibitem [{\citenamefont {Chiorescu}\ \emph {et~al.}(2010)\citenamefont
  {Chiorescu}, \citenamefont {Groll}, \citenamefont {Bertaina}, \citenamefont
  {Mori},\ and\ \citenamefont {Miyashita}}]{chiorescu2010magnetic}%
  \BibitemOpen
  \bibfield  {author} {\bibinfo {author} {\bibfnamefont {I.}~\bibnamefont
  {Chiorescu}}, \bibinfo {author} {\bibfnamefont {N.}~\bibnamefont {Groll}},
  \bibinfo {author} {\bibfnamefont {S.}~\bibnamefont {Bertaina}}, \bibinfo
  {author} {\bibfnamefont {T.}~\bibnamefont {Mori}},\ and\ \bibinfo {author}
  {\bibfnamefont {S.}~\bibnamefont {Miyashita}},\ }\bibfield  {title} {\bibinfo
  {title} {Magnetic strong coupling in a spin-photon system and transition to
  classical regime},\ }\href@noop {} {\bibfield  {journal} {\bibinfo  {journal}
  {Phys. Rev. B}\ }\textbf {\bibinfo {volume} {82}},\ \bibinfo {pages} {024413}
  (\bibinfo {year} {2010})}\BibitemShut {NoStop}%
\bibitem [{\citenamefont {Wu}\ \emph {et~al.}(2020)\citenamefont {Wu},
  \citenamefont {Xie}, \citenamefont {Ng}, \citenamefont {Mehanna},
  \citenamefont {Li}, \citenamefont {Attwood},\ and\ \citenamefont
  {Oxborrow}}]{HaoWu2020_cont}%
  \BibitemOpen
  \bibfield  {author} {\bibinfo {author} {\bibfnamefont {H.}~\bibnamefont
  {Wu}}, \bibinfo {author} {\bibfnamefont {X.}~\bibnamefont {Xie}}, \bibinfo
  {author} {\bibfnamefont {W.}~\bibnamefont {Ng}}, \bibinfo {author}
  {\bibfnamefont {S.}~\bibnamefont {Mehanna}}, \bibinfo {author} {\bibfnamefont
  {Y.}~\bibnamefont {Li}}, \bibinfo {author} {\bibfnamefont {M.}~\bibnamefont
  {Attwood}},\ and\ \bibinfo {author} {\bibfnamefont {M.}~\bibnamefont
  {Oxborrow}},\ }\bibfield  {title} {\bibinfo {title} {Room-temperature
  quasi-continuous-wave pentacene maser pumped by an invasive
  $\mathrm{Ce}:\mathrm{YAG}$ luminescent concentrator},\ }\href
  {https://doi.org/10.1103/PhysRevApplied.14.064017} {\bibfield  {journal}
  {\bibinfo  {journal} {Phys. Rev. Applied}\ }\textbf {\bibinfo {volume}
  {14}},\ \bibinfo {pages} {064017} (\bibinfo {year} {2020})}\BibitemShut
  {NoStop}%
\bibitem [{\citenamefont {Sloop}\ \emph {et~al.}(1981)\citenamefont {Sloop},
  \citenamefont {Yu}, \citenamefont {Lin},\ and\ \citenamefont
  {Weissman}}]{Sloop1981}%
  \BibitemOpen
  \bibfield  {author} {\bibinfo {author} {\bibfnamefont {D.~J.}\ \bibnamefont
  {Sloop}}, \bibinfo {author} {\bibfnamefont {H.}~\bibnamefont {Yu}}, \bibinfo
  {author} {\bibfnamefont {T.}~\bibnamefont {Lin}},\ and\ \bibinfo {author}
  {\bibfnamefont {S.~I.}\ \bibnamefont {Weissman}},\ }\bibfield  {title}
  {\bibinfo {title} {Electron spin echoes of a photoexcited triplet: Pentacene
  in p‐terphenyl crystals},\ }\href {https://doi.org/10.1063/1.442520}
  {\bibfield  {journal} {\bibinfo  {journal} {J. Chem. Phys.}\ }\textbf
  {\bibinfo {volume} {75}},\ \bibinfo {pages} {3746} (\bibinfo {year}
  {1981})}\BibitemShut {NoStop}%
\bibitem [{\citenamefont {Ai}\ \emph {et~al.}(2017)\citenamefont {Ai},
  \citenamefont {Chen}, \citenamefont {Feng},\ and\ \citenamefont
  {Xu}}]{Ai2017}%
  \BibitemOpen
  \bibfield  {author} {\bibinfo {author} {\bibfnamefont {Q.}~\bibnamefont
  {Ai}}, \bibinfo {author} {\bibfnamefont {P.}~\bibnamefont {Chen}}, \bibinfo
  {author} {\bibfnamefont {Y.}~\bibnamefont {Feng}},\ and\ \bibinfo {author}
  {\bibfnamefont {Y.}~\bibnamefont {Xu}},\ }\bibfield  {title} {\bibinfo
  {title} {Growth of pentacene-doped p-terphenyl crystals by vertical bridgman
  technique and doping effect on their characterization},\ }\href
  {https://doi.org/10.1021/acs.cgd.6b01900} {\bibfield  {journal} {\bibinfo
  {journal} {Cryst. Growth Des.}\ }\textbf {\bibinfo {volume} {17}},\ \bibinfo
  {pages} {2473} (\bibinfo {year} {2017})}\BibitemShut {NoStop}%
\end{thebibliography}%


\begin{thebibliography}{3}%
\makeatletter
\providecommand \@ifxundefined [1]{%
 \@ifx{#1\undefined}
}%
\providecommand \@ifnum [1]{%
 \ifnum #1\expandafter \@firstoftwo
 \else \expandafter \@secondoftwo
 \fi
}%
\providecommand \@ifx [1]{%
 \ifx #1\expandafter \@firstoftwo
 \else \expandafter \@secondoftwo
 \fi
}%
\providecommand \natexlab [1]{#1}%
\providecommand \enquote  [1]{``#1''}%
\providecommand \bibnamefont  [1]{#1}%
\providecommand \bibfnamefont [1]{#1}%
\providecommand \citenamefont [1]{#1}%
\providecommand \href@noop [0]{\@secondoftwo}%
\providecommand \href [0]{\begingroup \@sanitize@url \@href}%
\providecommand \@href[1]{\@@startlink{#1}\@@href}%
\providecommand \@@href[1]{\endgroup#1\@@endlink}%
\providecommand \@sanitize@url [0]{\catcode `\\12\catcode `\$12\catcode
  `\&12\catcode `\#12\catcode `\^12\catcode `\_12\catcode `\%12\relax}%
\providecommand \@@startlink[1]{}%
\providecommand \@@endlink[0]{}%
\providecommand \url  [0]{\begingroup\@sanitize@url \@url }%
\providecommand \@url [1]{\endgroup\@href {#1}{\urlprefix }}%
\providecommand \urlprefix  [0]{URL }%
\providecommand \Eprint [0]{\href }%
\providecommand \doibase [0]{https://doi.org/}%
\providecommand \selectlanguage [0]{\@gobble}%
\providecommand \bibinfo  [0]{\@secondoftwo}%
\providecommand \bibfield  [0]{\@secondoftwo}%
\providecommand \translation [1]{[#1]}%
\providecommand \BibitemOpen [0]{}%
\providecommand \bibitemStop [0]{}%
\providecommand \bibitemNoStop [0]{.\EOS\space}%
\providecommand \EOS [0]{\spacefactor3000\relax}%
\providecommand \BibitemShut  [1]{\csname bibitem#1\endcsname}%
\let\auto@bib@innerbib\@empty
\bibitem [{\citenamefont {Kajfez}(1995)}]{kajfez1995}%
  \BibitemOpen
  \bibfield  {author} {\bibinfo {author} {\bibfnamefont {D.}~\bibnamefont
  {Kajfez}},\ }\bibfield  {title} {\bibinfo {title} {Q-factor measurement with
  a scalar network analyser},\ }\href
  {https://digital-library.theiet.org/content/journals/10.1049/ip-map_19952142}
  {\bibfield  {journal} {\bibinfo  {journal} {IEE Proceedings - Microwaves,
  Antennas and Propagation}\ }\textbf {\bibinfo {volume} {142}},\ \bibinfo
  {pages} {369} (\bibinfo {year} {1995})}\BibitemShut {NoStop}%
\bibitem [{\citenamefont {Kajfez}()}]{kajfezwebsite}%
  \BibitemOpen
  \bibfield  {author} {\bibinfo {author} {\bibfnamefont {D.}~\bibnamefont
  {Kajfez}},\ }\href@noop {} {\bibinfo {title} {Q factor measurements, analog
  and digital}},\ \bibinfo {howpublished} {Available:
  https://people.engineering.olemiss.edu/darko-kajfez/assets/rfqmeas2b.pdf},\
  \bibinfo {note} {(Accessed: 2022-07-30)}\BibitemShut {NoStop}%
\bibitem [{\citenamefont {Breeze}\ \emph {et~al.}(2017)\citenamefont {Breeze},
  \citenamefont {Salvadori}, \citenamefont {Sathian}, \citenamefont {Alford},\
  and\ \citenamefont {Kay}}]{Breeze2017cqed}%
  \BibitemOpen
  \bibfield  {author} {\bibinfo {author} {\bibfnamefont {J.~D.}\ \bibnamefont
  {Breeze}}, \bibinfo {author} {\bibfnamefont {E.}~\bibnamefont {Salvadori}},
  \bibinfo {author} {\bibfnamefont {J.}~\bibnamefont {Sathian}}, \bibinfo
  {author} {\bibfnamefont {N.~M.}\ \bibnamefont {Alford}},\ and\ \bibinfo
  {author} {\bibfnamefont {C.~W.~M.}\ \bibnamefont {Kay}},\ }\bibfield  {title}
  {\bibinfo {title} {Room-temperature cavity quantum electrodynamics with
  strongly coupled dicke states},\ }\href
  {https://doi.org/10.1038/s41534-017-0041-3} {\bibfield  {journal} {\bibinfo
  {journal} {npj Quantum Inf.}\ }\textbf {\bibinfo {volume} {3}},\ \bibinfo
  {pages} {40} (\bibinfo {year} {2017})}\BibitemShut {NoStop}%
\end{thebibliography}%

\newpage
\newpage

\end{document}


\title{Supplementary Information\\Move aside pentacene: Diazapentacene doped para-terphenyl as a zero-field room-temperature maser with strong coupling for cavity quantum electrodynamics}

\author{Wern Ng}
\email{wern.ng@imperial.ac.uk}
 \affiliation{Department of Materials, Imperial College London, South Kensington, SW7 2AZ London, United Kingdom}

 \author{Xiaotian Xu}
 \affiliation{Department of Materials, Imperial College London, South Kensington, SW7 2AZ London, United Kingdom}
 
  \author{Max Attwood}
 \affiliation{Department of Materials, Imperial College London, South Kensington, SW7 2AZ London, United Kingdom}
 
  \author{Hao Wu}
 \affiliation{Center for Quantum Technology Research and Key Laboratory of Advanced Optoelectronic Quantum Architecture and Measurements (MOE), School of Physics, Beijing Institute of Technology, Beijing 100081, China}
  \affiliation{Beijing Academy of Quantum Information Sciences, Beijing 100193, China}
 
 \author{Zhu Meng}
 \affiliation{Department of Chemistry and Centre for Processible Electronics, Imperial College London, W12 0BZ London, United Kingdom}

 \author{Xi Chen}
  \affiliation{Department of Materials, Imperial College London, South Kensington, SW7 2AZ London, United Kingdom}
 \affiliation{Department of Computer Science, University of Southern California,  Los Angeles, California, USA}

 \author{Mark Oxborrow}
 \altaffiliation{m.oxborrow@imperial.ac.uk}
 \affiliation{Department of Materials, Imperial College London, South Kensington, SW7 2AZ London, United Kingdom}

\date{\today}

\maketitle

\section{Resonator coupling coefficients and $Q$-factor}
\begin{figure}[hbp]
  \includegraphics[width=0.9\linewidth]{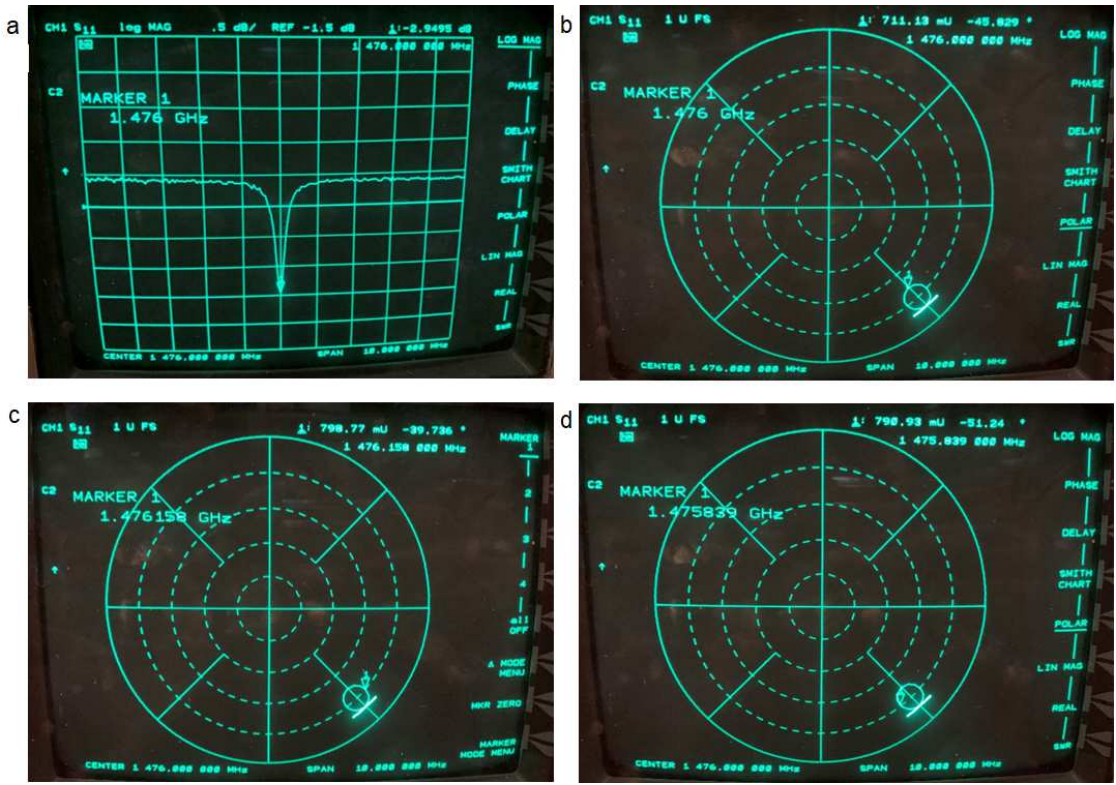}
  \caption{(a) $S_{11}$ dip of the main coupling loop (Port 1). (b), (c), (d) show the $Q$-circle of Port 1 on the $S_{11}$ polar plot of the VNA, with different marker positions.}
  \label{fig:maincoupling}
\end{figure}

The coupling coefficient of the main coupling loop (Port 1 in the main text) can be calculated using plots of the reflection coefficient ($\Gamma$) with frequency on the polar plot of a VNA (using `$Q$-circles'\cite{kajfez1995}). A detailed explanation of this technique is shown in the literature\cite{kajfez1995,kajfezwebsite}, but some working will be given. If the coupling was not lossy, a `$Q$-circle' as shown in Figure S~\ref{fig:maincoupling}(b)-(d) would have been intersecting with the boundary of the polar plot ($ |\Gamma| =1$), and the coupling coefficient would be $K=d/(2-d)$, with $d$ being the diameter of the $Q$-circle (which from Figure S~\ref{fig:maincoupling} is measured to be $d=0.16$). However, as seen in Figure S~\ref{fig:maincoupling} the coupling is lossy and the circle is detached from the polar plot boundary.



\begin{figure}[htbp]
\centering
  \includegraphics[width=0.5\linewidth]{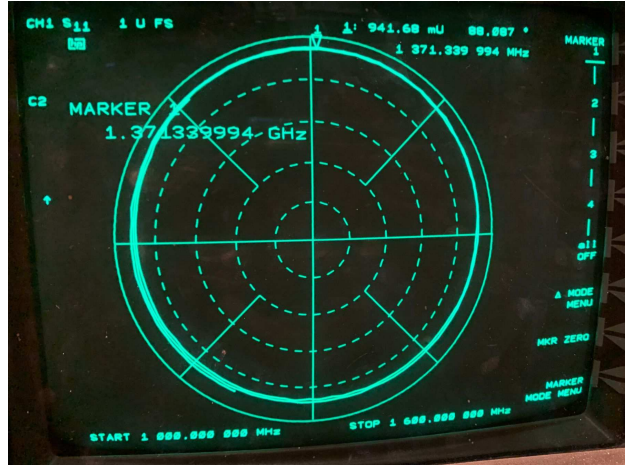}
  \caption{Auxiliary circle of Port 1, encompassing the smaller $Q$-circle seen in Figure S~\ref{fig:maincoupling}}
  \label{fig:auxcircle}
\end{figure}

\begin{figure}[htbp]
  \includegraphics[width=\linewidth]{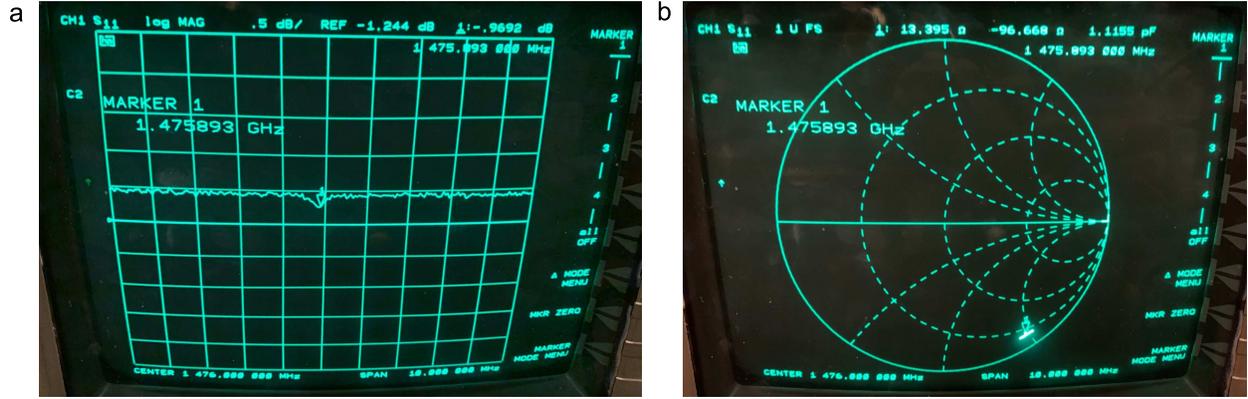}
  \caption{(a) $S_{11}$ dip of Port 2, which is severely undercoupled. (b) $Q$-circle of the weaker coupling loop, which has a negligible diameter. Though the plot is in Smith Chart form, the negligible diameter should be similar as in the polar plot.}
  \label{fig:weakport}
\end{figure}

\begin{figure}[htbp]
\centering
  \includegraphics[width=0.6\linewidth]{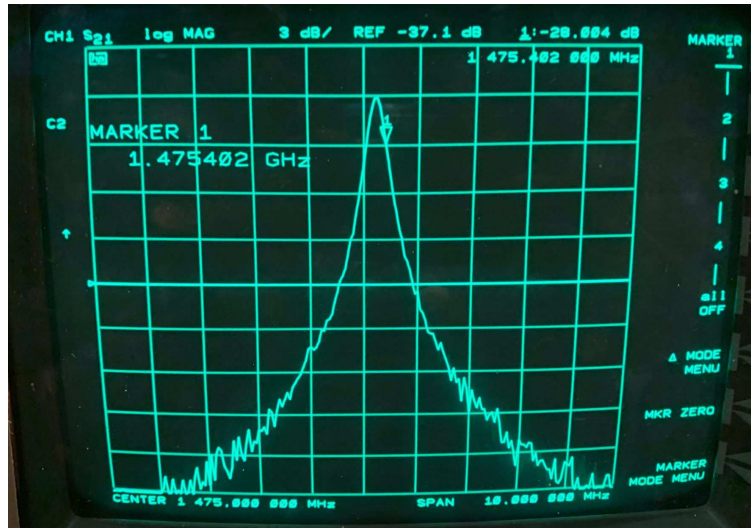}
  \caption{$S_{21}$ transmission of the cavity}
  \label{fig:s21}
\end{figure}

In this case, there should be another `auxiliary circle' which is ideally tangential to the $Q$-circle and the boundary of the polar plot. Then, the coupling coefficient can be calculated using the diameter of the auxiliary circle $d_2$ in the following way:
\begin{equation}
\label{eq:coupling}
    K=\frac{d}{d_2-1}
\end{equation}
This auxiliary circle was found in Figure S~\ref{fig:auxcircle}, and was assumed to be close enough to the polar plot boundary to be considered attached to it (plotting the auxiliary circle required re-calibrating the VNA at a larger frequency span). Its diameter was measured to be $d_2=1.81$ in the polar plot. Hence, from Equation~\ref{eq:coupling}, $K=0.20$ for Port 1. Though this working was shown for the frequency of 1.476 GHz, the coupling of Port 1 was virtually identical when tuned between 1.45 and 1.478 GHz for Pc and DAP resonances respectively. The second coupling loop (Port 2) was made to be extremely undercoupled (its loop is very small and it is spaced very far from the STO ring) so as to not lower the loaded $Q$ ($Q_L$). Figure S~\ref{fig:weakport} shows the $S_{11}$ dip and $Q$-circle of Port 2, where its $Q$-circle diameter $d$ is so small as to be negligible.
The $Q_L$ of the resonator can be calculated using the method outlined in Ref~\cite{kajfezwebsite} using the $Q$-circle measured through $S_{11}$ in Figure S~\ref{fig:maincoupling}, which gives $Q_L=1.476/(1.4762-1.4758)=3690$ (it was assumed lossy coupling does not affect the calculation). Alternatively, an $S_{21}$ measurement was also performed in Figure S~\ref{fig:s21}, where the negligible coupling of Port 2 would mean that $Q_L$ was only affected by Port 1. Measurement of the bandwidth of the peak gives $Q_L=3688$, which is virtually identical to that calculated from the $S_{11}$ $Q$-circle measurement. The unloaded $Q$ ($Q_u$) can be calculated if $Q_L$ and the coupling of the two ports are known:
\begin{equation}
\label{eq:qu}
    Q_L=\frac{Q_u}{1+K_1+K_2}
\end{equation}
Where $K_1$ and $K_2$ are the coupling coefficients of each port. Since $K_1=0.20$ and $K_2\approx0$,
then from Equation~\ref{eq:qu}, $Q_u=4428$.

The importance of these parameters is the following; if one wants to couple to the resonator to see a masing signal, then the coupling attained here ($K_1=0.20$) could be a good baseline to start with (the authors stress that the coupling doesn't need to be \textit{exactly} this amount). Alternatively, following the coupling shown by the $S_{11}$ of Figure S~\ref{fig:maincoupling}(a) would also be a useful guide.

\section{cQED Differential Equations and Parameters}
The set of coupled differential equations used for simulating Figure 6(c)-(f) in the main text are written as:
\begin{align}
    \frac{d}{dt}\langle a^\dagger a\rangle&= -\kappa_c\langle a^\dagger a\rangle+\kappa_c\Bar{n}+ig_e\left( \langle \Tilde{S}^+ a\rangle-\langle \Tilde{S}^+ a\rangle^*\right)\\
    \frac{d}{dt}\langle \Tilde{S}^+ a\rangle&=-\frac{1}{2}\left(\kappa_c+\gamma+\kappa_s+2i\Delta\right)\langle \Tilde{S}^+ a\rangle\\
    &\quad-ig_e\left(\frac{\langle \Tilde{S}^z\rangle+1}{2}+(1-1/N)\langle  \Tilde{S}^+ \Tilde{S}^-\rangle+\langle a^\dagger a\rangle\langle \Tilde{S}^z\rangle\right)\nonumber \\
    \frac{d}{dt}\langle \Tilde{S}^z\rangle&=-\gamma\langle \Tilde{S}^z\rangle-\frac{2ig_e}{N}\left( \langle \Tilde{S}^+ a\rangle-\langle \Tilde{S}^+ a\rangle^*\right)\\
    \frac{d}{dt}\langle  \Tilde{S}^+ \Tilde{S}^-\rangle&=-(\gamma+\kappa_s)\langle  \Tilde{S}^+ \Tilde{S}^-\rangle+ig_e\langle \Tilde{S}^z\rangle\left( \langle \Tilde{S}^+ a\rangle-\langle \Tilde{S}^+ a\rangle^*\right)
\end{align}
The derivation for these equations is given in previous work\cite{Breeze2017cqed}, but some definitions will be repeated here for convenience. The differential equations express the time evolution of the expectation values of four quantities: the cavity photon number $\langle a^\dagger a\rangle$, the spin-photon coherence $\langle a^\dagger \Tilde{S}^-\rangle=\langle \Tilde{S}^+ a\rangle^*$ (this substitution was made into the equations), the inversion $\langle \Tilde{S}^z\rangle$ and spin-spin correlation $\langle \Tilde{S}^+\Tilde{S}^-\rangle$ (before normalisation). $\kappa_c=2/T_2$ is the spin dephasing rate which is related to the spin-spin relaxation time $T_2$ of DAP:PTP. $\Bar{n}=(e^{hf/k_BT}-1)^{-1}$ is the average thermal photon population in the cavity at resonant frequency $f$ (evaluated to be 4097 at 290 K). $\gamma$ is the spin-lattice relaxation rate of DAP:PTP. $\Delta$ is the frequency detuning parameter. Finally, $N$ is the number of spins in DAP:PTP participating in the dynamics. $g_e$ and $\kappa_c$ have already been defined in the main text and in literature\cite{Breeze2017cqed}.

The initial conditions at $t=0$ were as follows: $\langle a^\dagger a\rangle_{t=0}=\Bar{n}$, $\langle \Tilde{S}^z\rangle_{t=0}=0.52$ since this is the ratio between the difference and sum of the populations ($(0.6-0.19)/(0.6+0.19)$), and $\langle \Tilde{S}^+ a\rangle_{t=0}=\langle  \Tilde{S}^+ \Tilde{S}^-\rangle_{t=0}=0$. It was assumed that there was no frequency detuning ($\Delta=0$), and that $\gamma=0.2$~MHz which was the upper limit on the decay time of the spins attained from ZF-trEPR in the main paper. Using the Runge-Kutta solver, the ODEs could be solved with the following remaining parameters giving the best fit to the experimental plot of $\langle a^\dagger a\rangle$; $\kappa_c=2\pi\times0.29$ MHz, $g_e=2\pi\times2.3$ MHz and $N=9.7\times10^{14}$. Once these parameters were found, the other three expectation values could be plotted in Figure 6 of the main text, where the plot of $\langle a^\dagger \Tilde{S}^-\rangle$ was obtained by taking the complex conjugate of $\langle \Tilde{S}^+ a\rangle$ (we note that $\langle a^\dagger \Tilde{S}^-\rangle$ was fully complex), and the plot of $\langle  \Tilde{S}^+ \Tilde{S}^-\rangle/N$ could be obtained using the value found for $N$. 

\section{DFT spin-orbit coupling matrix elements}
Table S~\ref{tab:dft} shows the spin-orbit coupling matrix that indicates a weak degree of spin-orbit coupling between singlet and triplet states in DAP:PTP. 

\def\arraystretch{0.8}%
\begin{table}[h!]
    \caption{Calculated spin-orbit coupling matrix elements between triplet states, $i$, and singlet states, $j$, of DAP. Units in cm$^{-1}$. }
\begin{ruledtabular}
    \begin{tabular}{ccccc} 
Triplet State, $i$ & Singlet State, $j$ & X & Y & Z \\
    \hline
    0 & 0 & 0.00 & 0.00 & -0.00 \\
    0 & 1 & 0.00 & 0.00 & -0.00 \\
    0 & 2 & -0.00 & -0.00 & -0.00 \\
    0 & 3 & -3.15 & -3.09 & 10.87 \\
    0 & 4 & 0.02 & -0.02 & 0.00 \\
    0 & 5 & -0.00 & -0.00 & 0.00 \\
    1 & 0 & 0.04 & -0.03 & 0.01 \\
    1 & 1 & 0.00 & 0.00 & 0.00 \\
    1 & 2 & 0.00 & 0.00 & -0.00 \\
    1 & 3 & 0.00 & -0.00 & -0.00 \\
    1 & 4 & 0.00 & 0.00 & -0.00 \\
    1 & 5 & -0.00 & -0.00 & 0.00 \\
    2 & 0 & 0.00 & 0.00 & 0.00 \\
    2 & 1 & 2.27 & 2.22 & -7.83 \\
    2 & 2 & -0.00 & 0.00 & 0.00 \\
    2 & 3 & 0.00 & 0.00 & 0.00 \\
    2 & 4 & -0.40 & -0.72 & -0.32 \\
    2 & 5 & 0.00 & -0.00 & -0.00 \\
    3 & 0 & -0.00 & 0.00 & 0.00 \\
    3 & 1 & 0.01 & -0.01 & 0.00 \\
    3 & 2 & -0.00 & -0.00 & 0.00 \\
    3 & 3 & 0.54 & 0.99 & 0.44 \\
    3 & 4 & -0.00 & -0.00 & -0.00 \\
    3 & 5 & -0.00 & -0.00 & -0.00 
    \end{tabular}
    \label{tab:dft}
\end{ruledtabular}
\end{table}
\newpage

\bibliography{supp}